\documentclass[5p,twocolumn]{elsarticle}

\UseRawInputEncoding
\usepackage{siunitx}
\usepackage{amsmath}
\usepackage{graphicx}
\usepackage{subfigure}
\usepackage{enumerate}
\usepackage{float}
\usepackage[percent]{overpic}
\usepackage{sidecap}
\usepackage{tipa}
\usepackage{epstopdf}
\usepackage{booktabs}
\usepackage{natbib}
\usepackage{hyperref}
\usepackage{lineno}
\modulolinenumbers[5]
\usepackage{appendix}
\usepackage{arydshln}
\usepackage{xcolor}
\usepackage{lipsum}

\DeclareSIUnit\atm{atm}
\DeclareSIUnit\calorie{cal}

\journal{Journal of Biomechanics}

\usepackage{numcompress}\biboptions{authoryear}

\begin{document}

\begin{frontmatter}

\title{Aerodynamic effects and performance improvements of running \\ in drafting formations}



\author[mainaddress]{Lukas Schickhofer\corref{correspondingauthor}}
\address[mainaddress]{KTM E-TECHNOLOGIES, Salzburg, AT-5081, Austria}
\cortext[correspondingauthor]{Corresponding author:}
\ead{lukas.schickhofer@ktm.com}

\author[secondaryaddress]{Henry Hanson}
\address[secondaryaddress]{ADIDAS, Herzogenaurach, DE-91074, Germany}

\begin{abstract}
Drafting as a process to reduce drag and to benefit from the presence of other competitors
is applied in various sports with several recent examples of competitive running in formations.
In this study, the aerodynamics of a realistic model of a female runner is calculated 
by computational fluid dynamics (CFD) simulations
at four running speeds of $\SI{15}{\kilo\meter\per\hour}$, $\SI{18}{\kilo\meter\per\hour}$, $\SI{21}{\kilo\meter\per\hour}$, and $\SI{36}{\kilo\meter\per\hour}$.
Aerodynamic power fractions of the total energy expenditure are found to be in the range of $2.6\%$-$8.5\%$.
Additionally, four exemplary formations are analysed with respect to their drafting potential
and resulting drag values are compared for the main runner and her pacers.
The best of the formations achieves a total drag reduction on the main runner of $75.6\%$.
Moreover, there are large variations in the drag reduction between the considered formations
of up to $42\%$ with respect to the baseline single-runner case. 
We conclude that major drag reduction of more than $70\%$ can already be achieved with fairly simple formations,
while certain factors, such as runners on the sides, can have a detrimental effect on drag reduction
due to local acceleration of the passing flow.
Using an empirical model for mechanical power output during running, 
gains of metabolic power and performance predictions are evaluated for all considered formations.
Improvements in running economy are up to $3.5\%$ for the best formation, leading 
to velocity gains of $2.3\%$. 
This translates to $\SI{154}{\second}$ ($\approx\SI{2.6}{\minute}$) saved over a marathon distance.
Consequently, direct conclusions are drawn from the obtained data for ideal drafting of long-distance running in highly packed formations.
\end{abstract}


\end{frontmatter}


\section{\textbf{Introduction}}
\label{Sec:1}

In most sports involving the race against other participants for time,
drag plays a crucial role.
It is defined as the aerodynamic force acting in an axial direction on
the athlete as he or she moves against the surrounding fluid, such as water or air.
As a result, it has become common practice both in sports and motorsports to 
benefit from the presence of other athletes or vehicles to alleviate the resistance of the fluid
by drafting in the associated wake region (i.e. \emph{slipstreaming}). 
In the wake there is not just drastically lower fluid velocity, but also a negative 
pressure coefficient leading to suction, which additionally benefits the trailing competitor.

While in motorsports with its high speeds and considerable research efforts,
drafting has long been investigated (e.g. \citet{katz1995race, romberg1971aerodynamics}), it is less researched in other sports. 
However, \citet{rundell1996effects} investigated drafting during speed skating and
showed large positive effects on metabolic activity, heart rate, and lactate response.
Also in swimming, drafting was associated with considerable energy saving and 
the optimal drafting distance was established at $0$-$\SI{0.5}{\meter}$ behind the leading swimmer,
causing power savings of up to $31\%$ \citep{bassett1991metabolic, chatard2003drafting}.
Furthermore, cycling is a sport of large benefits through aerodynamic drafting and riding in
formations, such as tightly packed pelotons \citep{edwards2007aerodynamic, blocken2018aerodynamic, malizia2020bicycle}. 
Apart from testing in wind tunnels, typically applying similarity principles through scaled 
models at similar Reynolds numbers, computational fluid dynamics (CFD) has become 
a widely used method to accurately assess drag distributions across athletes. 
\citet{blocken2013cfd} examined the effect of two drafting cyclists at various upper-body 
positions through CFD and found drag reductions of up 
to $27.1\%$ for the trailing cyclist, while also the leading cyclist's drag was decreased by up to $2.6\%$.
Moreover, drag reductions of $42\%$ to $48\%$ have been measured for drafting during cycling in the velodrome \citep{fitton2018impact}.
\citet{fuss2018slipstreaming} recently suggested an analytical model for slipstreaming in gravity-powered sports
such as ski cross and found significant advantages in gained velocity and glide distance with respect to a leading skier,
especially in a tucked body position.

Although running happens typically at fairly low speeds,
drafting shows also here a measurable impact on aerodynamic forces.
In a major study on the subject, \citet{pugh1970oxygen} demonstrated drag savings of up 
to $80\%$  when running at a middle-distance speed of $\SI{4.46}{\meter\per\second}$ against a slight head wind.
More recently, particularly in light of attempts to break the $2$-hour barrier of the marathon distance (e.g. \emph{INEOS 1:59 Challenge}, \emph{Nike Breaking2}),
increasing research efforts were directed to minimize drag for a main runner in formations.
\citet{hoogkamer2018modeling} used a reduced-order model to establish a sustainable velocity
of $\SI{5.93}{\meter\per\second}$ using cooperative drafting in a four-runner team.
\citet{polidori2020numerical} applied CFD to compute the drag and power savings of Kenenisa Bekele,
while running the Berlin marathon and using cooperative drafting.
It was shown that up to $57.3\%$ of aerodynamic power were gained in the ideal case of running directly
behind a front pacer at a distance of $\SI{1.3}{\meter}$ when compared to the case of running alone. 

The current study aims at shedding more light on the aerodynamic effects of running alone and in formations
by using a realistic model of a female runner and a validated CFD methodology.
Four different formations are chosen to reflect various running scenarios and their implications on the 
drag on the main runner as well as her pacers are discussed. 
The effect of the resulting drag reduction on the metabolic rate and running economy
is computed by using a mechanical power model of running.
As a result, possible performance predictions are made in terms of improvements of
running speed and time.
Finally, by applying the conclusions from this study, direct implications for an optimized drafting strategy are achieved.

\section{\textbf{Methods}}
\label{Sec:2}

\subsection{\textbf{Numerical method}}
\label{Sec:2A}

For the computation of the flow field variables and the resulting aerodynamic 
forces on the runner, computational fluid dynamics (CFD) using the finite-volume method is applied.
The Navier-Stokes equations describing the conservation of momentum are
discretized in its Reynolds-averaged (RANS) form.

Turbulence models have an important effect on aerodynamic force computations, mainly because of their modelling of momentum transfer near surfaces and in boundary layers. This in turn influences the locations of flow separation, which is crucial for correct drag predictions.
Here, turbulence is treated with the $k$-$\omega$ model in its shear-stress-transport (SST) formulation according to \citet{menter1994two}.
The model has been repeatedly shown to be suitable for external aerodynamics
and to give superior performance to the Spalart-Allmaras and the $k$-$\epsilon$ models
in flow scenarios with adverse pressure gradients and free shear layers, 
which do occur in our setup \citep{menter1992improved,hellsten1998some}.
\citet{defraeye2010computational} tested the performance of various turbulence 
treatments such as the most common RANS models and large eddy simulation
with respect to drag predictions for cycling aerodynamics.
They found the $k$-$\omega$ SST model to perform best in comparison
with experimental data (i.e. force and moment areas) from wind tunnel measurements,
with average discrepancies of $6\%$ and consistently below $11\%$.

\begin{figure*}[htbp]
\begin{centering}
\begin{overpic}[width=0.90\textwidth]{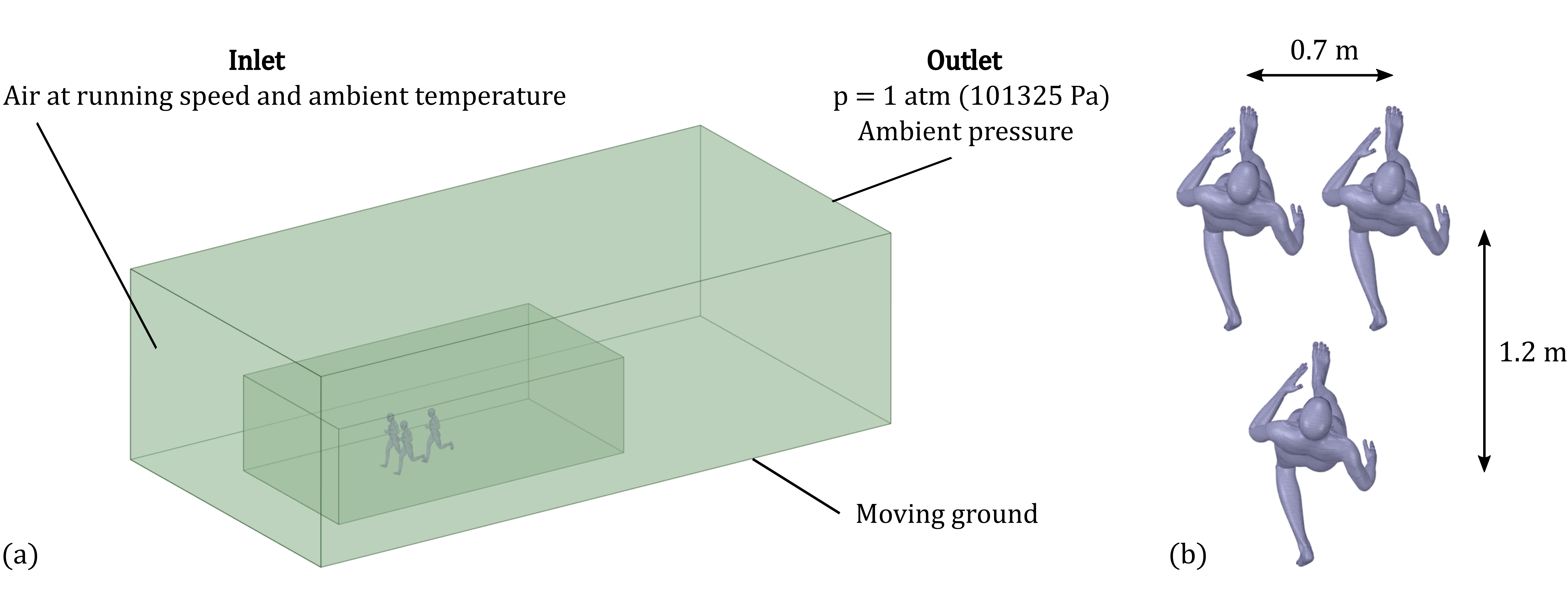}
\end{overpic}
\par
\end{centering}
\caption{Geometry of a formation of runners surrounded by the computational domain
and the imposed boundary conditions (a). The axial spacing of $1.2\,m$ and the lateral spacing 
of $0.7\,m$ between the runners are indicated (b).}
\label{Fig:2A-1} 
\end{figure*}

Moreover, \citet{crouch2016dynamic} has shown for cycling aerodynamics that the effect of 
the dynamical motion of the limbs on instantaneous drag is only minor.
Additionally, wind tunnel experiments by \citet{inoue2016wind} revealed an approximately 
$10\%$ increase in drag values of solo running when using a moving-belt system as compared with a stationary setup.
In their measurements the formation of unrealistic ground boundary layers was thus inhibited, as it is the case in this study.
We use a stationary model of an athlete together with a moving ground boundary,
which was found to be sufficient for the prediction of the global aerodynamic forces.

Furthermore, discretization in this study is performed with a second-order scheme
for the spatial derivatives. Pressure and velocity are calculated in a coupled approach.
The fluid is defined as air at ambient temperature of $T=\SI{293.15}{K}=\SI{20}{\celsius}$ 
and a kinematic viscosity of $\nu=\SI{1.516e-5}{\meter\squared\per\second}$. 
Additionally, the velocity inlet of the domain pictured in Fig. \ref{Fig:2A-1}(a)
imposes an axial flow velocity at the chosen running speeds, while the pressure outlet 
applies an ambient pressure of $p= \SI{101325}{\pascal}=\SI{1}{\atm}$.
The ground is defined as wall moving at the running speed, which ensures 
that no boundary layers form at the bottom that could invalidate the results.
A Dirichlet boundary condition, $u_{i}=0$, is applied for
all components of the flow velocity directly at the ground.
The surrounding walls are symmetry boundaries with frictionless properties.
This is achieved through a Neumann boundary condition $\partial u_{i} / \partial x_{i}=0$.

Tab. \ref{Tab:2A-1} gives the various speeds of this study with the respective Reynolds numbers 
of the flow, using the runner's height for the definition of the characteristic length.
While all four speeds are considered for the assessment of single-runner aerodynamics (cf. Sec. \ref{Sec:3A}),
the drafting formations are studied at a running speed of $\SI{21}{\kilo\meter\per\hour}$ (cf. Sec. \ref{Sec:3B}-\ref{Sec:3D}).
This is a high speed occurring during elite marathon running and allows for good comparison between
the considered formations. 
In fact, the current official marathon world record for men is 2:01:39 by Eliud Kipchoge
(September 16, 2018 during the Berlin Marathon; at an average speed of $\sim \SI{21}{\kilo\meter\per\hour}$) 
and for women is 2:14:04 by Brigid Kosgei 
(October 13, 2019 during the Chicago Marathon; at an average speed of $\sim \SI{19}{\kilo\meter\per\hour}$).
Here, we chose a running velocity at the upper range of these numbers to assess the ideal potential 
of drafting with respect to metabolic quantities, drafted velocity, and time savings.

\begin{table}[htbp]
\caption{Running speeds and associated Reynolds numbers $Re=\left( U_{\infty} L \right)/\nu$
at a characteristic length of $L=\SI{1.65}{\meter}$.}
\centering
\begin{tabular}{@{}llll@{}}
\toprule
Running speed ($\SI{}{\kilo\meter\per\hour}$) & Reynolds number \\
\midrule
$15$ ($\SI{4.16}{\meter\per\second}$) & $\SI{452770}{}$ \\
$18$ ($\SI{5}{\meter\per\second}$) & $\SI{544195}{}$ \\
$21$ ($\SI{5.83}{\meter\per\second}$) & $\SI{634531}{}$ \\
$36$ ($\SI{10}{\meter\per\second}$) & $\SI{1088390}{}$ \\
\bottomrule
\end{tabular}
\label{Tab:2A-1}
\end{table}

The inherent challenge of simulations lies in ensuring that physical effects 
are computed as accurately as possible.
Here, that means that flow separation lines and points, as well as boundary layer thickness and characteristics 
are captured for the investigated Reynolds numbers, which lie in the turbulent regime (cf. Tab. \ref{Tab:2A-1}).
In order to avoid inaccurate prediction of aerodynamic forces on athletes at high Reynolds numbers, 
as described by \citet{meile2006aerodynamics}, it is crucial to optimize near-surface grid resolutions,
such that no numerically induced jumps of separation points occur.
This is important, as an artificially delayed flow separation would lead to a sharp drop in the 
pressure drag, which is the dominant component of total drag of bluff bodies (such as the human body).
A thorough verification and validation of the applied numerical approach is presented in \ref{App:A}.
The solution of the discretized continuity and momentum equations is performed 
with the solver \emph{ANSYS Fluent 19.2}, post-processing of flow variables is achieved using
the post-processing tool \emph{ParaView} and drag distribution and 
resulting power values are calculated with the free scientific programming 
language \emph{GNU Octave}.

\subsection{\textbf{Geometry and mesh}}
\label{Sec:2B}

For the numerical setup we apply the realistic, three-dimensional  geometry of a female
runner with a height of $L=\SI{1.65}{\meter}$, a projected frontal area of $A\approx\SI{0.44}{\meter\squared}$,
and a body mass of $m\approx\SI{55}{\kilogram}$.
The geometry represents an average of 59 individual track athletes, scanned in-house and
positioned into a running pose. The obtained pose-invariant statistical model for the female human body 
was reconstructed from the high-quality scan data using the method outlined by \citet{colaianni2014pose}.
The axial and lateral spacing between the runners is chosen according to estimates 
of realistic running conditions. 
\citet{polidori2020numerical} for instance uses an axial distance of $\SI{1.3}{\meter}$ and 
a lateral shoulder-to-shoulder distance of $\SI{0.3}{\meter}$, which are taken from
an actual drafting situation during a marathon race.

The computational domain shown in Fig. \ref{Fig:2A-1}(a) has a distance of $\SI{7}{\meter}$
from the main runner to the upstream boundary and $\SI{13}{\meter}$ to the downstream boundary.
It contains far-field refinement as well as three refinement stages close to the runner's surfaces.
Furthermore, 10 prism layers (i.e. inflation layers) are inserted directly at all solid boundaries
with a steady growth factor of the prism layer thickness of $1.2$.
These measures ensure that the dimensionless wall distance of $y^{+}\approx1$ holds in the 
entire domain and all boundary layers are well resolved.
Due to the local mesh refinement the cell count varies sharply between formations depending
on the number of runners included. It rises from $40$ million cells for the single-runner case
up to $107$ million cells for the final formation including a total of four runners.
The near-surface mean cell size of $\SI{3.125e-3}{\meter}$ and the outer-boundary mean cell size of $\SI{0.196}{\meter}$
for the final grid settings have been chosen following a grid convergence study (cf. \ref{Sec:A1}).
Homogeneous, gradual cell growth is applied between the near and far field.
Locally, especially in the boundary layer regions, perpendicular mesh dimensions are significantly lower (more than by a factor 10)
than the near-surface mean cell size, which is the cell size of the finest grid level just outside the prism layers.
A depiction of the considered formations of this study is shown in Fig. \ref{Fig:3B-1}-\ref{Fig:3B-2}. 

\subsection{\textbf{Calculation of power values}}
\label{Sec:2C}

Various empirical models exist for the calculation of the mechanical power
output while running.
\citet{fukunaga1980effect} found through experiments on athletic runners
on force platforms a relation between running velocity $u$ and the 
exerted power for forward motion as 
\begin{equation}
P_{R,0} = 0.436u^{2.01},  
\label{Eq:2C-1} 
\end{equation}
with $u$ in $\SI{}{\meter\per\second}$ and $P_{R}$ in $\SI{}{\watt\per\kilogram}$.
In a study using motion tracking, \citet{cavagna1977mechanical} also
took into account additional energy expenditure created by the
movement of the limbs. They arrive at a function for the specific running power per unit mass as 
\begin{equation}
P_{R,0} = 9.42 + 4.73u + 0.266u^{1.993}.
\label{Eq:2C-2} 
\end{equation}
Here, $u$ is given in $\SI{}{\kilo\meter\per\hour}$ and $P_{R,0}$ in $\SI{}{\calorie\per\kilogram\per\minute}$.
For our considered cases, both approaches give similar results with
deviations in the range of only $2$-$5\%$, however, 
Eq. (\ref{Eq:2C-2}) is used in the following power calculations due to
being the more complete model of running motion.
By `more complete' we mean that it takes into account both internal and 
external work performed for the modelling of the total mechanical work
\begin{equation}
W_{tot} = W_{ext}+W_{int}.
\label{Eq:2C-2-1} 
\end{equation}
Here, external work $W_{ext}$ is due to acceleration and lift of the centre of mass,
while internal work refers to both the translational and rotational acceleration of the 
limbs relative to the trunk. 
This approach has also been suggested more recently by the work of 
\citet{saibene2003biomechanical}, \citet{pavei2019comprehensive}, and \citet{gray2020model}.

Using Eq. (\ref{Eq:2C-2}), we arrive at the power
\begin{equation}
P_{R} = P_{R,0} \cdot \frac{4.1868}{60} \cdot m
\label{Eq:2C-3} 
\end{equation}
in $W$ for a runner of mass $m$ in $\SI{}{\kilogram}$, in the following simply called \emph{running power}.
Furthermore, there is a certain aerodynamic power $P_{A}$, used
to overcome drag $F_{D}$, which is defined as
\begin{equation}
P_{A} = F_{D} \cdot u = \frac{1}{2} \rho u^{3} C_{D} A.
\label{Eq:2C-4} 
\end{equation}
Here, the expression for $F_{D}$ assumes stagnant air and the absence of crosswinds.
As a result, the total mechanical power generated while running
is the sum of the running power $P_{R}$ and the aerodynamic power $P_{A}$:
\begin{equation}
P_{tot} = P_{R}+P_{A}.
\label{Eq:2C-5} 
\end{equation}
Using the total mechanical power $P_{tot}$ and the associated efficiency $\epsilon$
at a certain run speed by \citet{cavagna1977mechanical},
the metabolic power
\begin{equation}
P_{meta} = \frac{P_{tot}}{\epsilon},
\label{Eq:2C-6} 
\end{equation}
which gives the metabolic energy expenditure during running, can be computed.

It must be emphasized here that the science behind efficiency of locomotion
and in particular of walking and running is far from concluded.
In both running and walking the muscles utilize energy stored during a previous phase of stretching
in the following phase of contraction.
As stated by \citet{cavagna1977mechanical}, the efficiency of walking reaches a maximum 
of $0.35$-$0.40$ at intermediate speeds due to the expected properties of the contractile component of muscle.
This maximum of efficiency follows the force-velocity relation and the trends due to initial efficiency of muscle \citep{hill1964efficiency}.
Also in cycling the efficiency of locomotion was found to reach maximum values at intermediate speeds ($\sim 0.22$ according to the early study of \citet{dickinson1929efficiency},
$\sim 0.25$ according to \citet{ettema2009efficiency}).
In running, however, the efficiency increases steadily with speed (from $0.45$ to $0.70$), suggesting that
the positive work during that activity derives mainly from the passive recoil of elastic muscle tissue and 
not from the active shortening of contractile muscle components \citep{cavagna1977mechanical, saunders2004factors}.
In support of this hypothesis, \citet{cavagna1968positive} found that the useful effect of pre-streched
muscle on the performed work increases with the speed of stretching and shortening.
Furthermore, \citet{komi1978utilization} investigated this utilization of elastic energy during 
eccentric-concentric contraction of muscles in human motion and found 
that up to $90\%$ of energy produced in the pre-stretching phase is recovered in case of 
counter-movement jumps.
More recently, \citet{komi2000stretch} used data based on in-vivo force measurements, buckle-transducer
technique, and optic-fiber technique to show that the stretch-shortening cycle in human skeletal muscle 
leads to considerable performance enhancement compared to simple concentric action.
With respect to running economy, \citet{hunter2015muscle} showed a clear positive correlation
of stretch-shortening cycle potentiation and increases of running economy 
and recommended eccentric force development (e.g. by resistance training) for performance improvements.
Nonetheless, we point out that efficiency and total mechanical power models 
are inherently difficult to define due to the sometimes conflicting understanding of internal and external 
work performed during human locomotion and the question, whether internal work should appear in
total mechanical power and efficiency expressions, as has been suggested by  
\citet{winter1979new} and more recently by \citet{minetti2001bipedalism}
(see e.g. \citet{ettema2009efficiency} for a discussion of these issues with regards to cycling,
and \citet{williams2000dynamics} with regards to running).

\section{\textbf{Results}}
\label{Sec:3}

\subsection{\textbf{Single-runner aerodynamics}}
\label{Sec:3A}

\begin{figure}[htbp]
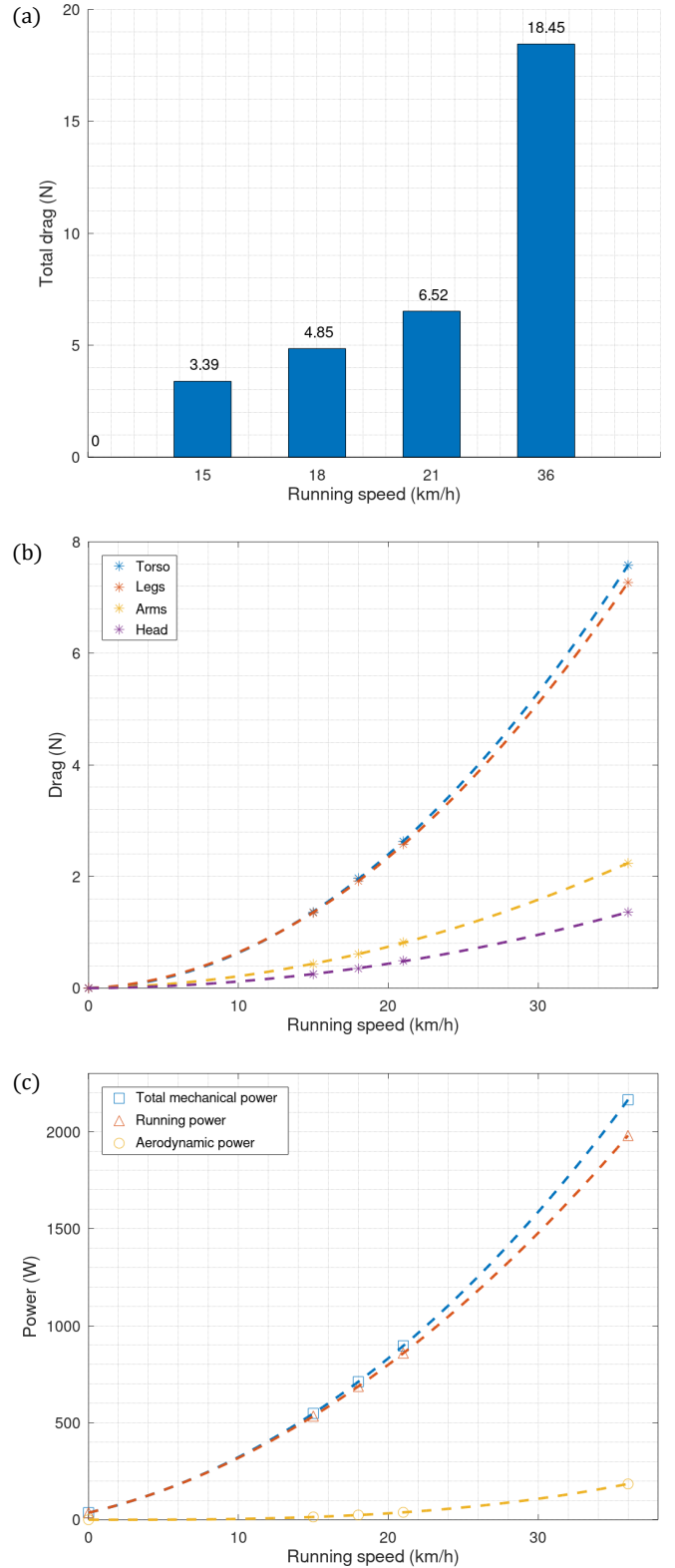

\begin{centering}
\begin{overpic}[width=0.50\textwidth]{Figure2.pdf}
\end{overpic}
\par
\end{centering}
\caption{Total drag over the range of considered running speeds (a), 
as well as drag acting on the body parts of the runner (b). 
Total mechanical power, consisting of running power computed according to the empirical
model by \citet{cavagna1977mechanical} and aerodynamic power, as a function of speed (c).
Polynomial functions of second order (aerodynamic forces and running power) and of third order (aerodynamic and total mechanical power)
are used for fitting the data.}
\label{Fig:3A-1} 
\end{figure}

Using the geometry of the single runner,
the drag forces acting on various body parts are computed over the range of relevant velocities (cf. Fig. \ref{Fig:3A-1}(b)).
It can be seen that the drag acting on torso and legs is highest, with both parts roughly affected
by the same aerodynamic loads. Each make for a fraction of $39$-$40\%$ of the total drag on the runner.
Furthermore, arms are affected considerably less, with a share of approximately $13\%$,
and the head with $8\%$ of the total drag.
These estimates give an overview of the sensitivity of the various parts of the 
body towards drag loads and are directly related to the frontal area of each part (cf. Eq. (\ref{Eq:2C-4})).
Fig. \ref{Fig:3A-1}(a) gives the total drag on the runner for the four simulated running speeds.
While the drag for lower speeds of $15$-$\SI{21}{\kilo\meter\per\hour}$, which are relevant for long distance 
disciplines like marathon, lies in the range of $3.39$-$\SI{6.52}{\newton}$, it sharply rises to $\SI{18.45}{\newton}$ 
at the higher considered speed of $\SI{36}{\kilo\meter\per\hour}$. This demonstrates the (quadratically) increasing relevance of
drag reduction at higher speeds.
Moreover, the associated power values are calculated using the definitions of Sec. \ref{Sec:2C}.
Fig. \ref{Fig:3A-1}(c) shows the share of aerodynamic and running power in the total mechanical power
over the considered range of velocity.
However, the actual energy per unit time, which needs to be made available during motion, is the 
metabolic power (cf. Eq. (\ref{Eq:2C-6})).

Here, the aerodynamic component provides a fraction of the metabolic power of
$2.6\%$, $3.4\%$, $4.3\%$, and $8.5\%$ for
$\SI{4.16}{\meter\per\second}$ ($\SI{15}{\kilo\meter\per\hour}$), $\SI{5}{\meter\per\second}$ ($\SI{18}{\kilo\meter\per\hour}$), $\SI{5.83}{\meter\per\second}$ 
($\SI{21}{\kilo\meter\per\hour}$), and $\SI{10}{\meter\per\second}$ ($\SI{36}{\kilo\meter\per\hour}$).
These can be considered as threshold values for the maximum savings in metabolic energy consumption
during drafted running. 

\begin{figure*}[htbp]
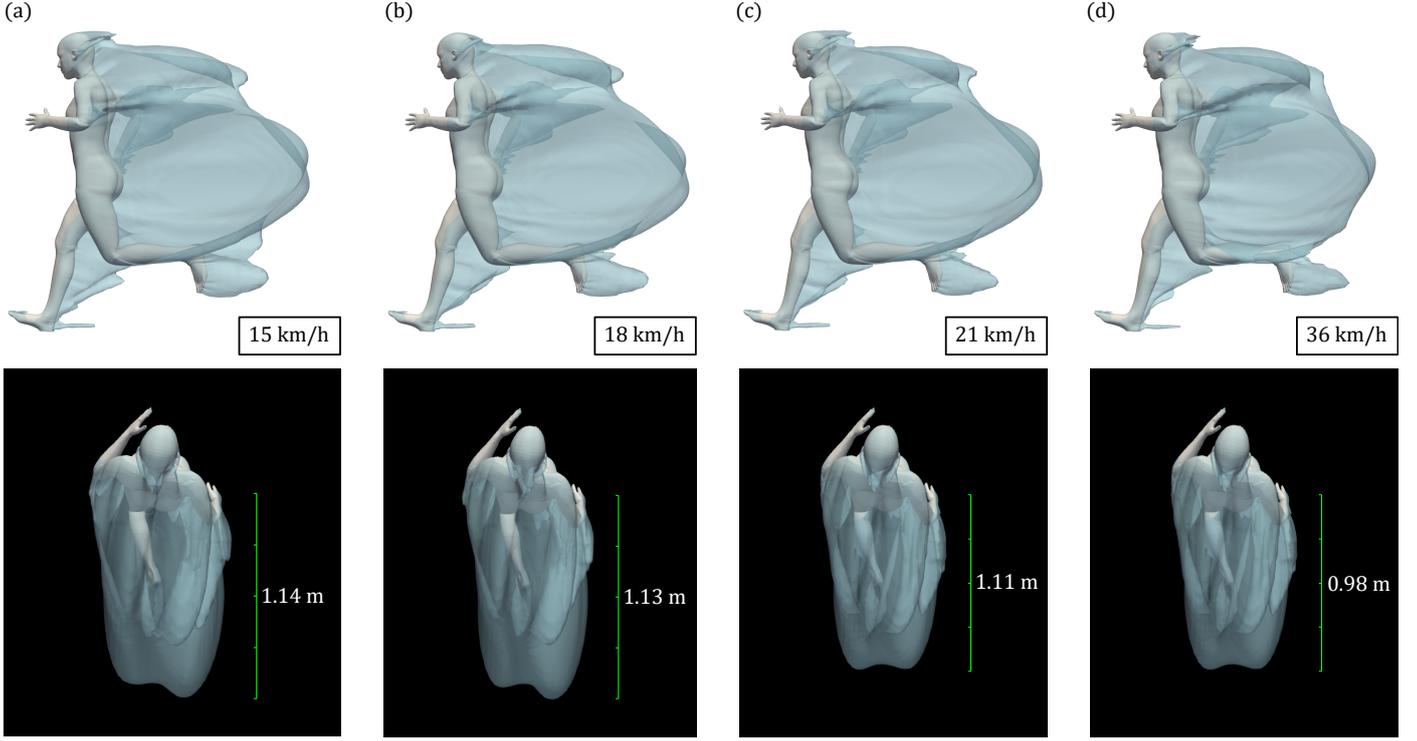

\begin{centering}
\vspace{0.2cm}
\begin{overpic}[width=1.00\textwidth]{Figure3.pdf}
\end{overpic}
\par
\end{centering}
\caption{Wake region behind the runner visualized by isosurfaces of total pressure at $p_{tot}=0$.}
\label{Fig:3A-2} 
\end{figure*}

The resulting wake regions are depicted in Fig. \ref{Fig:3A-2}. They are visualized by
isosurfaces of total pressure, giving the boundary between the negative pressure region
in the wake and the positive pressure region due to recovery of the flow.
There is a notable decrease of the wake length from $\SI{1.14}{\meter}$ at $\SI{15}{\kilo\meter\per\hour}$ to
$\SI{0.98}{\meter}$ at $\SI{36}{\kilo\meter\per\hour}$ due to increased turbulent mixing.
The wake length is a good indication of the most beneficial range for drafting,
as the suction provided by the negative total pressure in the wake slightly pulls a following runner.
However, even at a length larger than that there will be a considerable drag-reducing effect due to flow
deceleration of the head wind.

\subsection{\textbf{Drag comparison of formations}}
\label{Sec:3B}

\begin{figure}[htbp]
\begin{centering}
\begin{overpic}[width=0.50\textwidth]{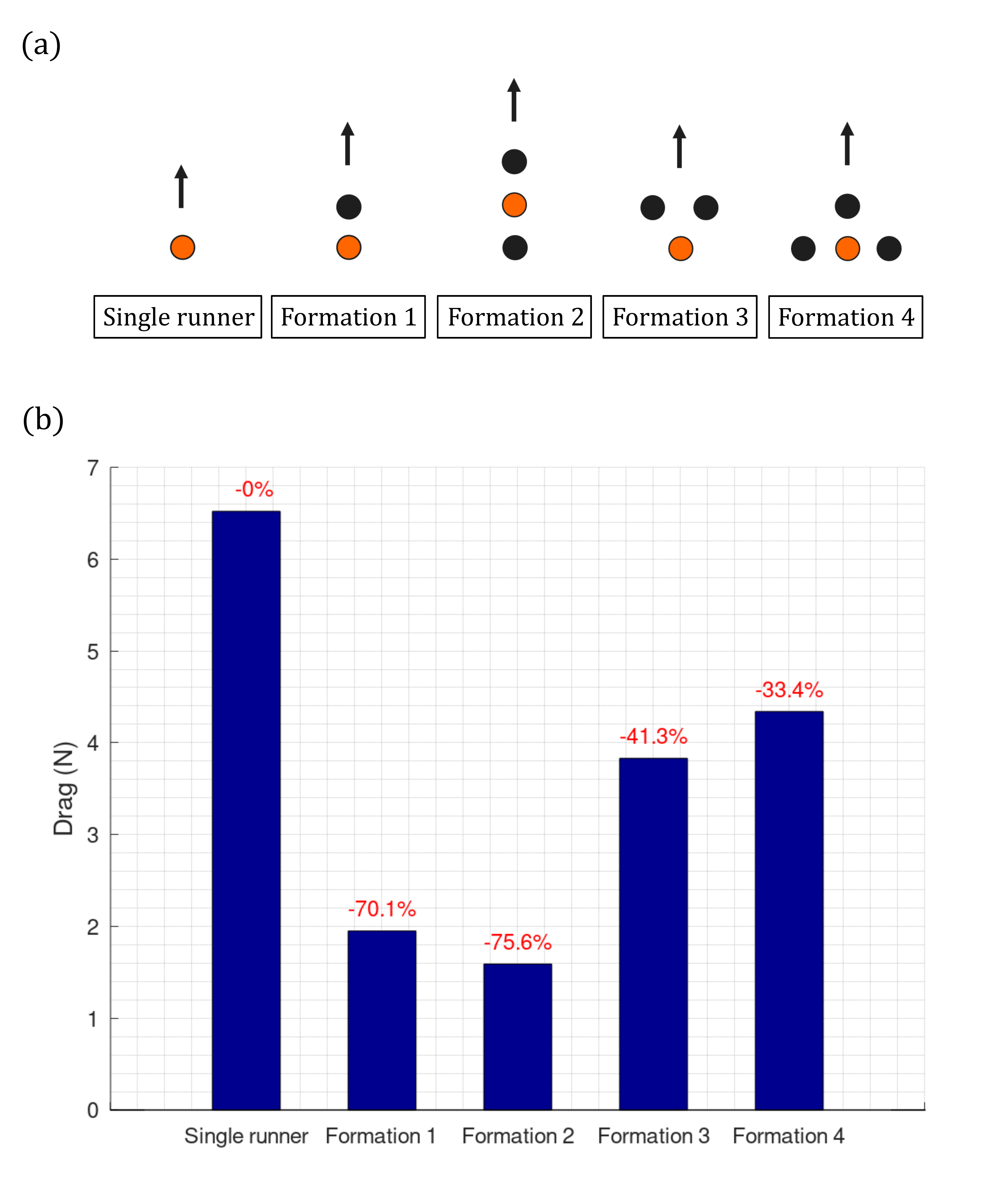}
\end{overpic}
\par
\end{centering}
\caption{Sketch of the investigated formations (a) together with their effect on the 
main runner's drag (b). The formations are compared to the case of the single runner.
Axial and lateral spacing between the runners follows the values introduced in Fig. \ref{Fig:2A-1}(b).}
\label{Fig:3B-1} 
\end{figure}

At first we consider the performance of the investigated formations with respect to
their effect on the main runner's drag at a running speed of $\SI{21}{\kilo\meter\per\hour}$.
It can be concluded from Fig. \ref{Fig:3B-1} that formation $2$ leads to the 
largest reduction of drag of $75.6\%$, followed by formation $1$ with $70.1\%$.
Formations $3$ and $4$ cause significantly less drag reduction of $41.3\%$ and $33.4\%$, respectively.

\begin{figure*}[htbp]
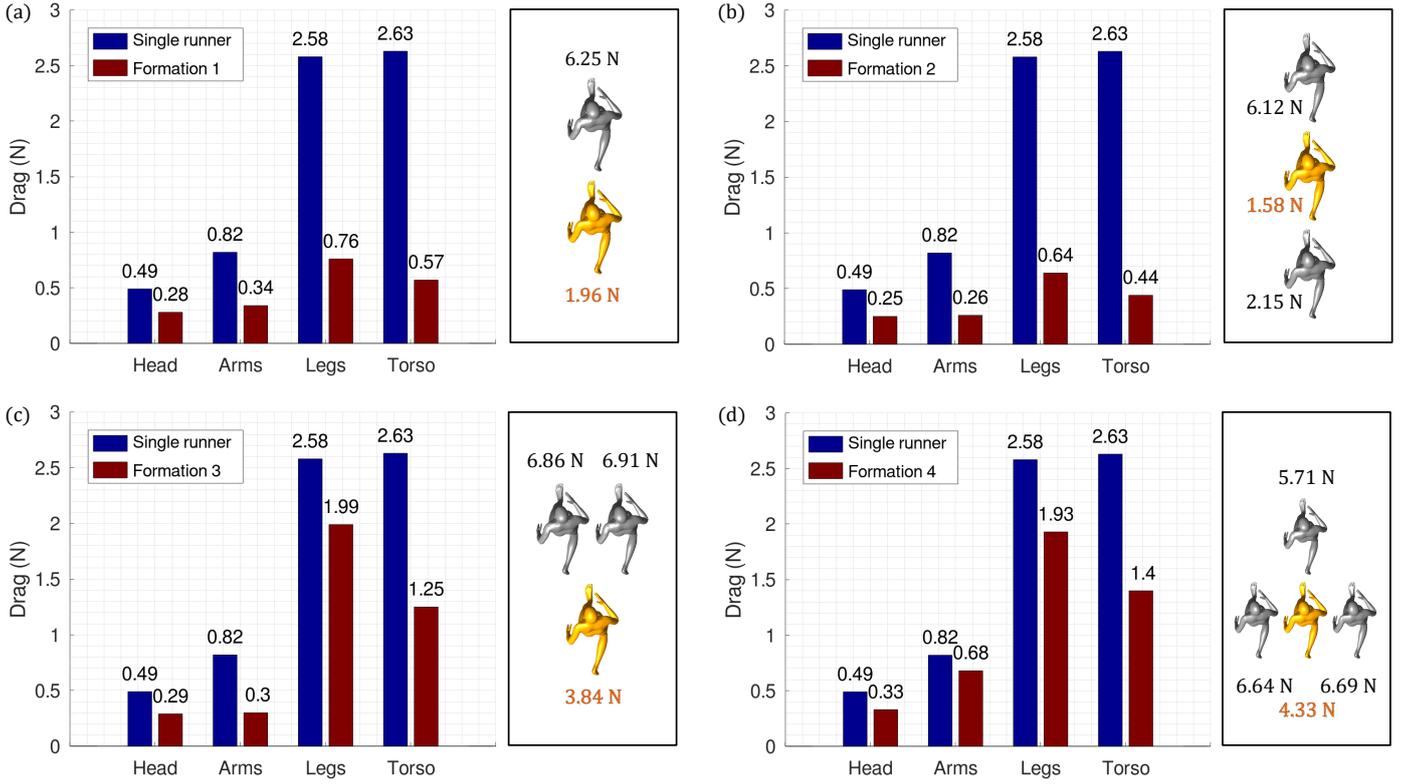

\begin{centering}
\begin{overpic}[width=1.00\textwidth]{Figure5.pdf}
\end{overpic}
\par
\end{centering}
\caption{Drag acting on the various body parts of the main runner alongside the total drag for 
all involved runners, including pacers, for the formations 1 (a), 2 (b), 3 (c), and 4 (d).}
\label{Fig:3B-2} 
\end{figure*}

Furthermore, considering the drag distribution across the whole body of the main runner,
it can be recognized that the dominant part of the drag acts on legs and torso (cf. Sec. \ref{Sec:3A}).
Thus, it is crucial for a drafting formation to reduce the drag in these areas.
Looking at Fig. \ref{Fig:3B-2}, it can be seen that formations $1$-$2$ achieve this, while 
formations $3$-$4$ cannot adequately shield the body of the main runner from oncoming air.
Additionally, it is shown that while the drag reduction on head and arms is fairly equal
for formations $1$-$3$, formation $4$ shows almost no improvement for the drag on the arms.
The reasons for the varying aerodynamic performances of formations $1$-$4$ are elaborated further in 
Sec. \ref{Sec:3C}. 

\begin{figure}[htbp]
\begin{centering}
\begin{overpic}[width=0.50\textwidth]{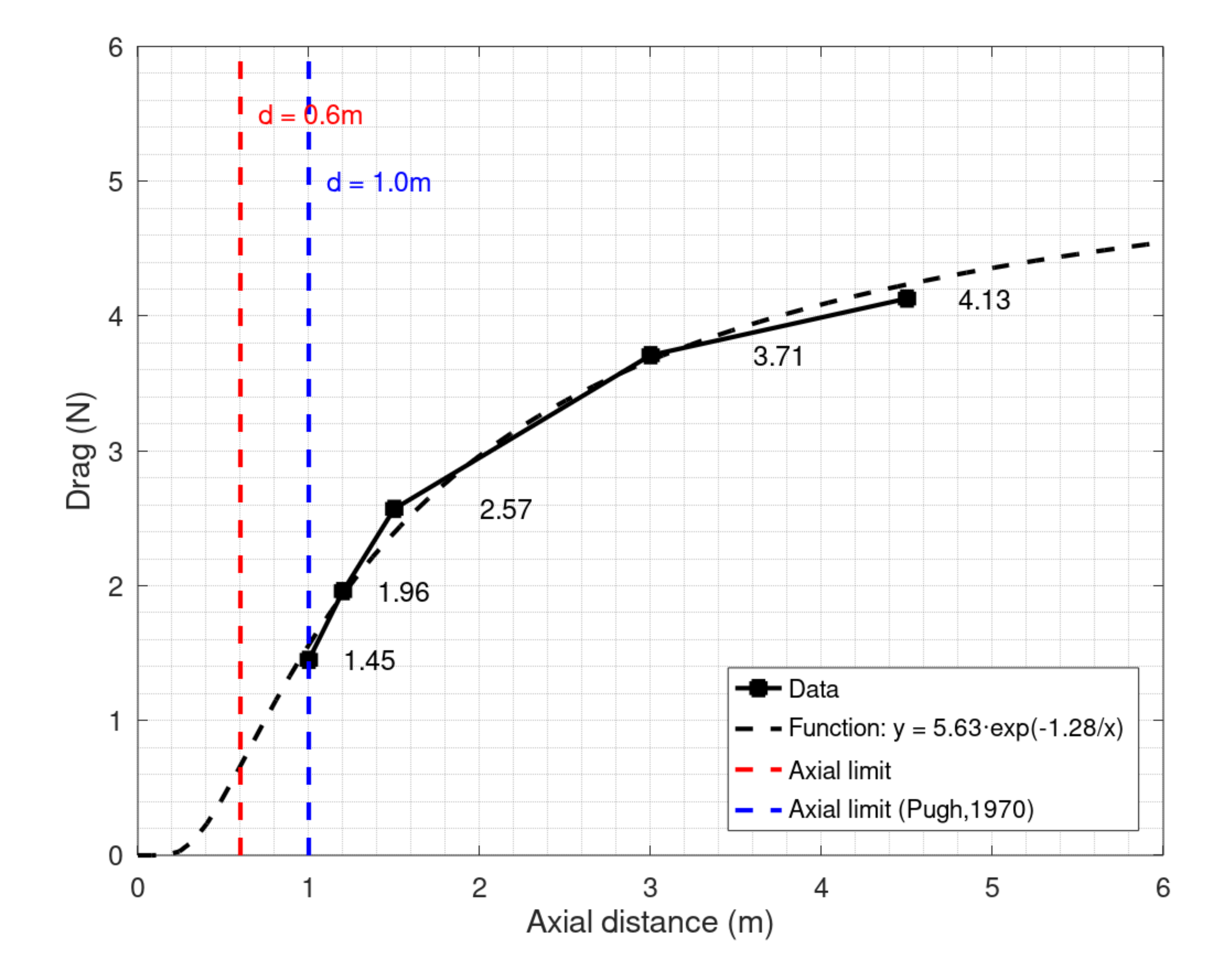}
\end{overpic}
\par
\end{centering}
\caption{Drag on the main runner as a function of the distance to the leading pacer in formation 1.
The numerical data is fitted using an exponential function with a determination coefficient of $R^{2}=0.99$.
In reality, there is an axial limit which prohibits the further approach of the main runner to the pacer.
This limit is estimated from geometric considerations as $\SI{0.6}{\meter}$. \citet{pugh1970oxygen} estimates
this minimum distance during drafting at $\SI{1.0}{\meter}$.}
\label{Fig:3B-3} 
\end{figure}

Using the results for formation $1$, the distance between main runner and pacer was further varied
to give the drag values shown in Fig. \ref{Fig:3B-3}.
An exponential function of the type $y = a + b \cdot \exp \left( -c / x \right)$ with $a = 0$, $b = 5.63$, and $c = 1.28$
was used to fit the data and has the desired property of asymptotically approaching zero drag at small distances 
and the single runner's maximum drag at large distances.
The values of the parameters were estimated by least-squares optimization using the Levenberg-Marquardt algorithm \citep{levenberg1944method,marquardt1963algorithm}.
The absolute axial limit of separation between the two runners, given the dimension and stride length of the applied geometry,
was found to be approximately $\SI{0.6}{\meter}$.
In reality, however, this limit will barely be reached and the runners will be unable to approach 
already considerably earlier.
\citet{pugh1970oxygen} estimates a natural limit of axial spacing of $\SI{1}{\meter}$ for middle-distance running
and states a possible drag reduction at this distance of up to $80\%$.
In this study we find a drag of $\SI{1.45}{\newton}$ acting on the main runner at a distance of $\SI{1}{\meter}$, which corresponds to $22.2\%$
of the drag on the single runner and thereby lies in good agreement with the results by \citet{pugh1970oxygen}.

\subsection{\textbf{Aerodynamic effects on runners within formations}}
\label{Sec:3C}

\begin{figure*}[htbp]
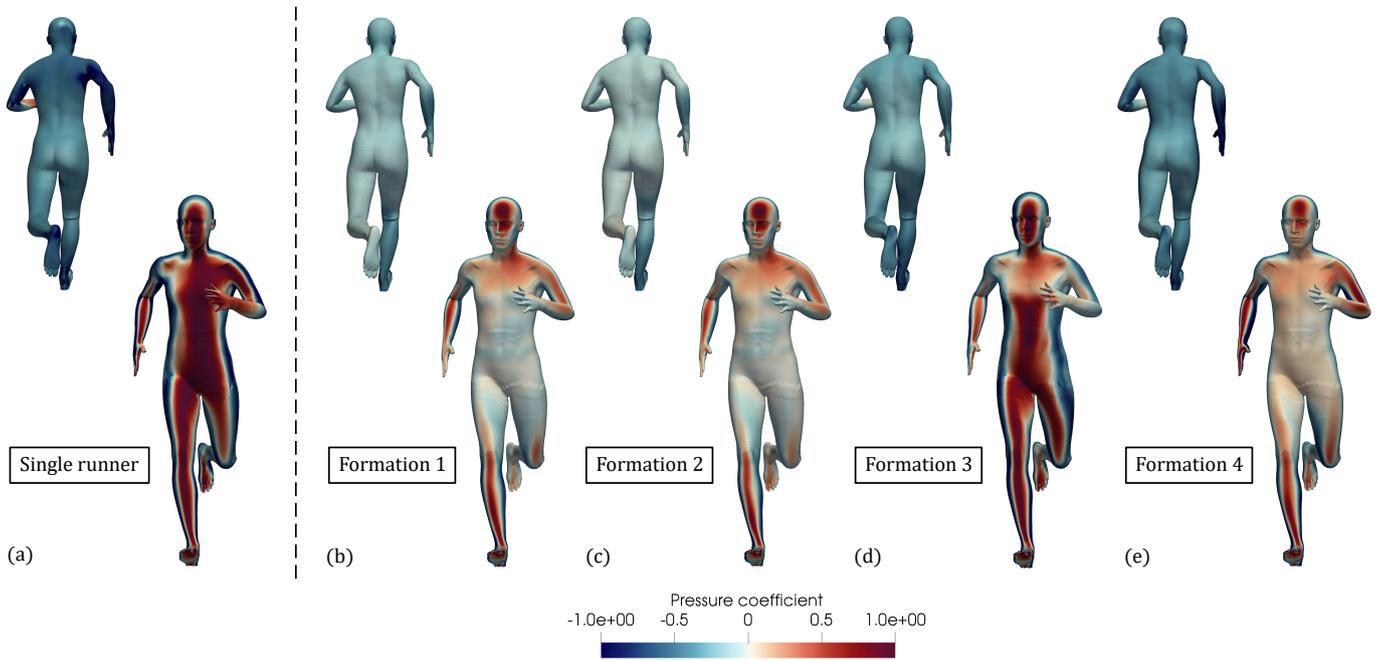

\begin{centering}
\begin{overpic}[width=1.00\textwidth]{Figure7.pdf}
\end{overpic}
\par
\end{centering}
\caption{Pressure coefficient on the main runner for the single-runner case (a) compared to 
the formations 1-4 (b)-(e).}
\label{Fig:3C-1} 
\end{figure*}

Air resistance while running manifests itself in static pressure increase
as the air is being decelerated and diverted at the front of the runner.
Fig. \ref{Fig:3C-1} provides insight to the magnitude of the pressure acting
on the main runner for the various formations. The pressure coefficient
\begin{equation}
c_{p} = \frac{p-p_{\infty}}{\frac{1}{2} \rho U_{\infty}^{2}} = \frac{p-p_{\infty}}{p_{0}-p_{\infty}},  
\label{Eq:3B-1} 
\end{equation}
with the free-stream pressure of $p_{\infty}= \SI{101325}{\pascal} = \SI{1}{\atm}$
and a relative free-stream velocity of $U_{\infty} = \SI{5.83}{\meter\per\second} = \SI{21}{\kilo\meter\per\hour}$, 
indicates regions of high or low relative pressure.

\begin{figure*}[htbp]
\begin{centering}
\vspace{0.4cm}
\begin{overpic}[width=1.00\textwidth]{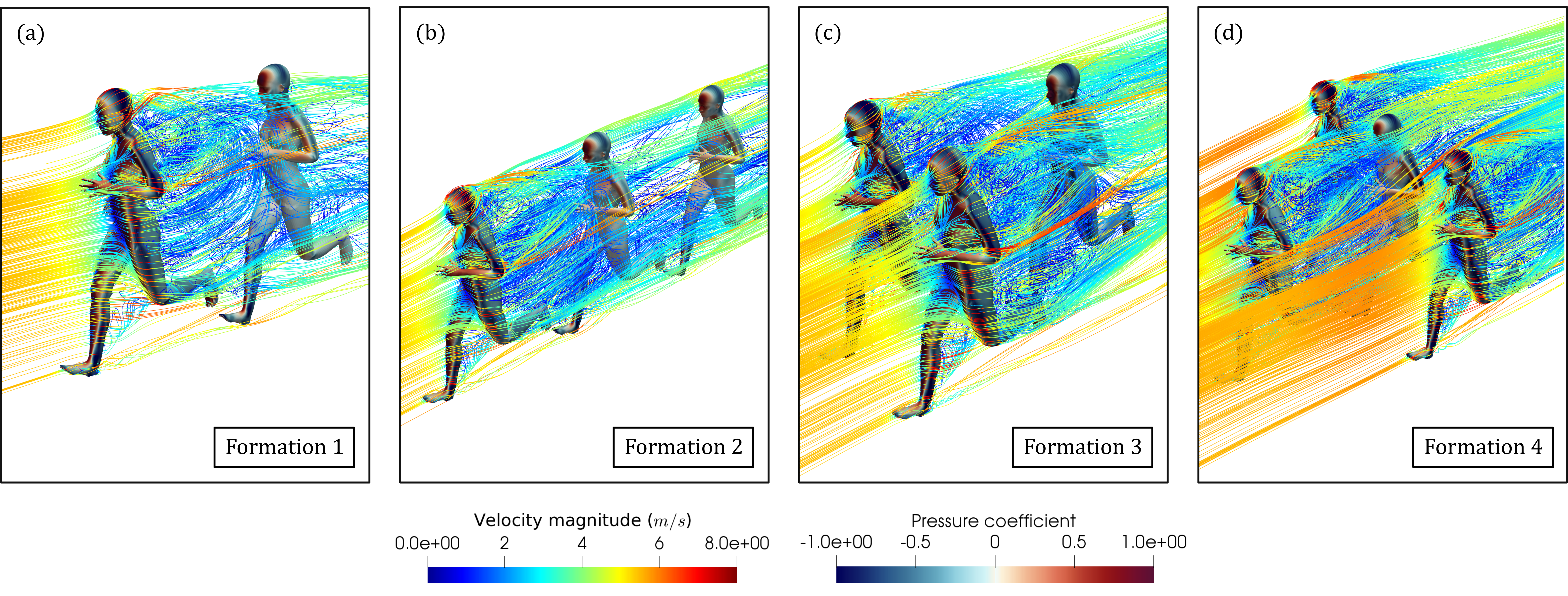}
\end{overpic}
\par
\end{centering}
\caption{Streamlines seeded at the front of the leading pacers showing the diversion
of air around the main runner for all four formations.
Streamlines are coloured by velocity magnitude and the runners' surfaces by pressure coefficient.}
\label{Fig:3C-2} 
\end{figure*}

\begin{figure*}[htbp]
\begin{centering}
\begin{overpic}[width=1.00\textwidth]{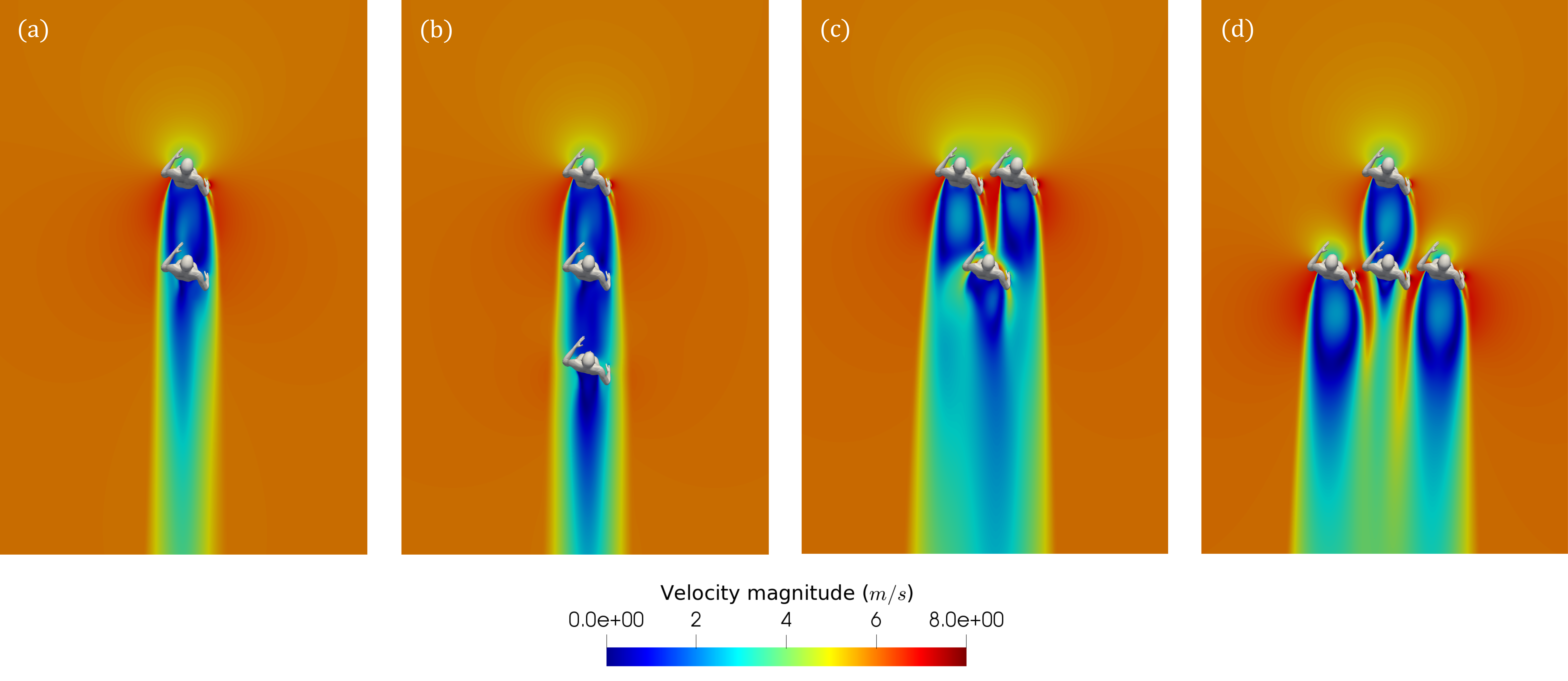}
\end{overpic}
\par
\end{centering}
\caption{Velocity magnitude contours in the horizontal plane section through the runners' centre.
Formations 1-4 are shown in images (a)-(d).}
\label{Fig:3C-3} 
\end{figure*}

Fig. \ref{Fig:3C-1} confirms the beneficial pressure distribution for formations $1$ and $2$.
This means that there are no large areas of high pressure coefficient at the front of the runner,
nor any large regions of low negative pressure coefficient at the back, which 
would be an indication for suction force slightly pulling the runner back.
Formation $3$ on the other hand shows a high positive pressure coefficient at the 
front, which already hints at the leakage of oncoming flow between the front pacers
that can be seen in Fig. \ref{Fig:3C-2}(c) and Fig. \ref{Fig:3C-3}(c).
While formation $4$ does not display any significantly high pressure peaks at the
front, it is the elevated negative pressure coefficient at the back that makes it less 
advantageous than formations $1$-$2$ (cf. Sec. \ref{Sec:3B}).

\begin{figure*}[htbp]
\begin{centering}
\begin{overpic}[width=0.75\textwidth]{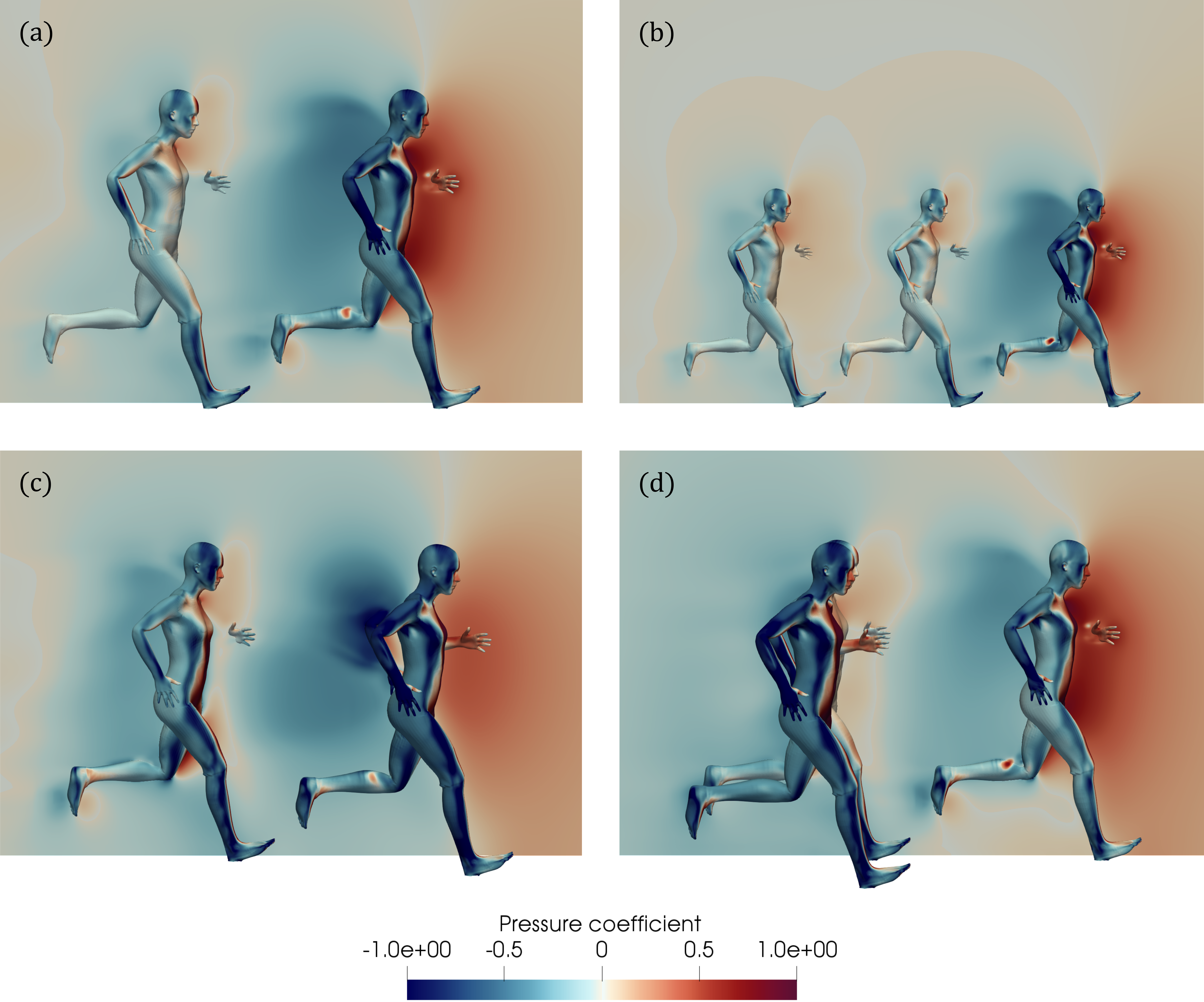}
\end{overpic}
\par
\end{centering}
\caption{Pressure coefficient in the vertical plane section through the main runner's centre.
Formations 1-4 are depicted in images (a)-(d).}
\label{Fig:3C-4} 
\end{figure*}

Fig. \ref{Fig:3C-2}-\ref{Fig:3C-3} both demonstrate the reasons for the large differences 
in aerodynamic performance between the formations.
In case of formations $1$ and $2$ the main runner is well embedded inside the wake region of 
the front pacer and therefore only affected by low-speed, almost stagnant air relative to herself.
Formation $2$ has the additional advantage of a rear pacer, whose high-pressure stagnation
region at the front positively affects the low-pressure wake region of the main runner by 
raising the pressure coefficient to a value closer to zero, thus lowering suction.
This can be seen when comparing Fig. \ref{Fig:3C-4}(a) and Fig. \ref{Fig:3C-4}(b).
Furthermore, Fig. \ref{Fig:3C-2}(c) and Fig. \ref{Fig:3C-3}(c) clearly demonstrate a major weakness 
of formation $3$, which is the passing air between the front pacers.
This leads to a large region of high stagnation pressure directly at the area of the impacting air 
on the main runner, which is also visible in Fig. \ref{Fig:3C-1}(d) and  Fig. \ref{Fig:3C-4}(c).
Also formation $4$ is suboptimal due to air being accelerated between the main runner and the 
pacers on the side. This leads to the aforementioned effect of higher drag at the arms 
compared to the other formations (cf. Sec. \ref{Sec:3B}).
Additionally, this effect of locally increased flow velocity between the runners leads to a shortening 
of the wake region of slow air for the main runner and thus negatively impacts her drafting conditions (cf. Fig. \ref{Fig:3C-3}(d)).

\subsection{\textbf{Metabolic power savings and performance predictions}}
\label{Sec:3D}

Using the drag values from Sec. \ref{Sec:3B}, we can compute the associated aerodynamic
power and the total mechanical power using Eq. (\ref{Eq:2C-2})-(\ref{Eq:2C-5}).
Tab. \ref{Tab:3D-1} lists the results for the single runner and the four formations.
Additionally, relative changes in aerodynamic power and total power are given by 
comparison with the results of the single-runner case.
The resulting metabolic power needed for running in the considered formations is computed by
using the efficiency of running at the given velocity as detailed in \citet{cavagna1977mechanical}.
Furthermore, running economy as the specific metabolic rate per unit mass is given together with its relative changes in Tab. \ref{Tab:3D-2}.
As described by \citet{kipp2019extrapolating}, the relationship between improvements in running economy
and improvements in velocity is not a linear, but rather a curvilinear one.
This means that for a given improvement in running economy, velocity gains will be smaller at higher speeds and larger at lower speeds.
At elite marathon speeds of around $~\SI{5.5}{\meter\per\second}$ there is roughly a $2/3^{\text{rds}}$ increase in velocity for each percent increase in running economy.
By applying the conversion factors of \citet{kipp2019extrapolating} we arrive at estimates for the velocity gains for each formation in Tab. \ref{Tab:3D-3}. The relative changes in speed are $2.1\%$ (formation 1), $2.3\%$ (formation 2), $1.3\%$ (formation 3), and $1.1\%$ (formation 4).
\begin{table*}[htbp]
\caption{Aerodynamic power $P_{A}$, and total mechanical power $P_{tot}$
computed with Eq. (\ref{Eq:2C-4}) and (\ref{Eq:2C-5}) and by computing the 
running power $P_{R}=\SI{858.1}{\watt}$ at $\SI{21}{\kilo\meter\per\hour}$ ($\SI{5.83}{\meter\per\second}$) using the model of \citet{cavagna1977mechanical}.
Savings $\Delta P_{A}$ and $\Delta P_{tot}$ are given with respect to the case of the runner alone.}
\centering
\begin{tabular}{@{}llllll@{}}
\toprule
Formation & $F_{D}$ ($\SI{}{\newton}$) & $P_{A}$ ($\SI{}{\watt}$) & $\Delta P_{A}$ & $P_{tot}$ ($\SI{}{\watt}$) & $\Delta P_{tot}$ \\
\midrule
Single runner & $6.52$ & $38.0$ & - & $896.1$ & - \\
$1$ & $1.96$ & $11.4$ & $70.0\%$ & $869.5$ & $3.0\%$ \\
$2$ & $1.58$ & $9.2$ & $75.8\%$ & $867.3$ & $3.2\%$ \\
$3$ & $3.84$ & $22.4$ & $41.1\%$ & $880.5$ & $1.7\%$ \\
$4$ & $4.33$ & $25.3$ & $33.4\%$ & $883.4$ & $1.4\%$\\
\bottomrule
\end{tabular}
\label{Tab:3D-1}
\end{table*}
\begin{table*}[htbp]
\caption{Metabolic power $P_{meta}$ according to Eq. (\ref{Eq:2C-6}) and running economy $RE$, as well as percent improvements of the 
running economy $\Delta RE$ for each of the formations with respect to the single-runner case.
Efficiency of locomotion for a running velocity of $\SI{21}{\kilo\meter\per\hour}$ ($\SI{5.83}{\meter\per\second}$) is taken as $\epsilon=0.64$ \citep{cavagna1977mechanical}.}
\centering
\begin{tabular}{@{}llll@{}}
\toprule
Formation & $P_{meta}$ ($\SI{}{\watt}$) & $RE$ ($\SI{}{\watt\per\kilogram}$) & $\Delta RE$ \\
\midrule
Single runner & $1400.2$ & $25.5$ & - \\
1 & $1358.6$ & $24.7$ & $3.1\%$ \\
2 & $1355.2$ & $24.6$ & $3.5\%$ \\
3 & $1375.8$ & $25.0$ & $2.0\%$ \\
4 & $1380.3$ & $25.1$ & $1.6\%$ \\
\bottomrule
\end{tabular}
\label{Tab:3D-2}
\end{table*}
\begin{table*}[htbp]
\caption{Drafted running speeds $u_{draft}$ for all formations alongside potential time $t_{draft}$
and time savings through drafting $\Delta t_{draft}$ over the marathon distance of $\SI{42195}{\meter}$ by comparison with the undrafted single-runner case.}
\centering
\begin{tabular}{@{}lllll@{}}
\toprule
Formation & $u_{draft}$ ($\SI{}{\meter\per\second}$) & $t_{draft}$ ($\SI{}{\second}$) & $\Delta t_{draft}$ ($\SI{}{\second}$) \\
\midrule
1 & $5.95$ & $7092$ ($\approx\SI{1.970}{\hour}$) & $142$ ($\approx\SI{2.4}{\minute}$) \\
2 & $5.96$ & $7080$ ($\approx\SI{1.967}{\hour}$) & $154$ ($\approx\SI{2.6}{\minute}$) \\
3 & $5.91$ & $7140$ ($\approx\SI{1.983}{\hour}$) & $98$ ($\approx\SI{1.6}{\minute}$) \\
4 & $5.89$ & $7164$ ($\approx\SI{1.990}{\hour}$) & $70$ ($\approx\SI{1.2}{\minute}$) \\
\bottomrule
\end{tabular}
\label{Tab:3D-3}
\end{table*}

\section{Discussion and conclusions}

Drafting and the resulting aerodynamic drag reduction cause a notable effect on running 
and the energy expenditure used for it.
This study shows the overall aerodynamics and associated energy expenditure during running
for a range of relevant speeds. Considerable shares of up to $8.5\%$ at a velocity of $\SI{36}{\kilo\meter\per\hour}$ are computed.

Additionally, differences in drag, pressure distribution across the runners,
and potential power savings are demonstrated for four formations that could arise during running competitions.
As shown in Sec. \ref{Sec:3B}, simple formations of axial spacing of the pacers 
(i.e. one pacer in front, as in formation 1, or one pacer in front and one in the back, as in formation 2) already lead to considerable 
reduction of over $70\%$ in drag at a distance of $\SI{1.2}{\meter}$. 
Furthermore, formations 1 and 2 also show a clear positive effect on leading and trailing 
pacers, since their overpressure and underpressure regions interact with the main runner.
This effect has also been noted by \citet{beaumont2019does} together with a measurable 
advantage for the pacers with respect to oxygen consumption.

We further deduce an empirical relationship of drag force acting on the main runner versus
her distance to a leading pacer, which can be used for future prediction of a pacer's aerodynamic effects (cf. Fig. \ref{Fig:3B-3}). 
The results are congruent with experimental data by \citet{pugh1970oxygen},
who predicts drag decrease of up to $80\%$ when running behind a pacer at $\SI{1}{\meter}$.
\citet{davies1980effects}, who claimed possible drag savings in the range of $80$-$85\%$,
also confirms these results.

Moreover, the drag distribution at various body parts is examined in Sec. \ref{Sec:3B}.
From this and the explanations in Sec. \ref{Sec:3C}, we can see a large sensitivity of the
drag acting on the main runner if pacers are positioned laterally, such as in formations $3$ and $4$.
In formation $3$, flow leaks between the front pacers, causing large pressure peaks at the runner's
centre (cf. Fig. \ref{Fig:3C-1}).
Formation $4$ on the other hand has the negative effect of pacers at the runner's
sides and the resulting acceleration of air between the runners leads to a relatively low
drag reduction of only $33.4\%$ compared to formation $1$ with $70.1\%$.
This suggests that runners at the side should be avoided if optimal drag savings are the goal.

In essence, there are three aerodynamic effects of drafting that lead to a reduction of
axial force (i.e. drag) while running:
\begin{enumerate}[(i)]
\item \emph{Shielding against fast, oncoming air.} \\
Therefore, there is no sharp deceleration of air and resulting formation of areas of high stagnation
pressure on the runner's surface.
\item \emph{Suction effects due to the region of negative total pressure in the leading runner's wake.} \\
Any object moving through air forms a wake region of relatively low pressure. 
This underpressure can be harnessed to the benefit of a trailing runner.
The wake region of negative total pressure has been found in this study to extend to about $\SI{1.14}{\meter}$ at $\SI{15}{\kilo\meter\per\hour}$ down to $\SI{0.98}{\meter}$ at 
$\SI{36}{\kilo\meter\per\hour}$ from the trailing runner's back.
\item \emph{Delay of separation points due to turbulent free shear layers.} \\
Turbulence levels are naturally increased in the high-shear-stress regions of the wake, 
where air is displaced against the stagnant surroundings.
This raises momentum transfer in the boundary layers of the runner and delays separation due
to formation of a turbulent boundary layer from a laminar one. As a result, pressure drag (i.e. form drag)
drops, which is the dominant source of drag of bluff bodies.
\end{enumerate}

Eventually, the aerodynamic power savings are computed in this study and listed together with the
total mechanical power and the metabolic power output in Sec. \ref{Sec:3D}.
The derived metabolic power and running economy of up to $\SI{1400.2}{\watt}$ and $\SI{25.5}{\watt\per\kilogram}$
for the single runner at a speed of $\SI{21}{\kilo\meter\per\hour}$ agrees well with values from literature for trained athletes \citep{kipp2017does, batliner2018does}.
The largest improvements of running economy occur for formation $1$ with $3.1\%$
and formation $2$ with $3.5\%$.
These metabolic gains are slightly higher than the $2.84\%$ gain computed
by \citet{polidori2020numerical} for the best position of Kenenisa Bekele's drafting strategy 
during the 2019 Berlin marathon. 
The reason for this is most likely the larger axial spacing of $\SI{1.3}{\meter}$ to the front pacer
as opposed to $\SI{1.2}{\meter}$ in the current study.
Furthermore, a particular benefit of this study is the additional performance prediction in
terms of improvements of velocity and time for the investigated formations.
By applying the curvilinear relationship between gains in velocity and running economy suggested by \citet{kipp2019extrapolating},
we deduce possible speed gains in the range of $1.1$-$2.3\%$.
Time savings of $\SI{1.2}{\minute}$ (formation 4) to $\SI{2.6}{\minute}$ (formation 2) are achieved over the marathon distance. 

In this study we focus on the impact of aerodynamic drafting on 
running economy. 
However, there are a multitude of other factors that influence running economy and performance.
With respect to biomechanics, some of these factors are (lower) limb mass distribution, Achilles tendon moment arm 
and lower body musculotendinous structures for reutilization of elastic energy,
as well as running style and gait patterns (e.g. so-called \emph{pose running} 
to benefit from elastic recoil forces in the lower limbs), among others \citep{williams1987relationship, saunders2004factors, barnes2015running}.

Concluding from this work, it is possible to give targeted suggestions for formation
patterns that are best suited for cooperative drafting.
It is shown that axially positioned pacers give the best performance and that laterally positioned pacers
should be running at a distance further away from the main runner to avoid local flow acceleration, while
still contributing positively to drag reduction.
Such an approach has been applied with the recent successful world record attempt by Eliud Kipchoge
during the \emph{INEOS 1:59 Challenge}, where a staggered formation in the shape of an inverted V has been chosen.
Future studies could explore a larger number of formations by CFD and subsequently
derive reduced-order models for parameters such as number of pacers, or axial and lateral spacing.
\bigskip

\noindent
\textbf{Acknowledgements} \\
This work was realised through research budget from ADIDAS and KTM E-TECHNOLOGIES.
Computational resources were provided by the parallel computing cluster at KTM in Mattighofen, Austria.
\bigskip

\noindent
\textbf{Conflict of interest statement} \\
The authors have nothing to disclose that would have biased this work.

\appendix
\section{Verification and validation}
\label{App:A}

Since errors and uncertainties are an unavoidable part of any simulation method such 
as CFD, measures must be taken to quantify their levels and to minimize them.
According to the guidelines by the \citet{american1998aiaa}, both verification and validation 
form a crucial first step of any CFD simulation.
While verification can be seen as the process of ensuring that `the equations are solved right',
validation means ensuring `that the right equations are solved' \citep{roache1998verification}. \\
In this work, we first perform a grid convergence study by successively refining the mesh 
until no significant changes in the relevant quantities are occurring at the chosen high-order schemes (i.e. verification).
Then we compare the obtained drag coefficients of typical bluff bodies to well-established values from literature.
Finally, since no wind tunnel measurement results are available for the specific runner geometry we use,
we compare the numerically computed drag coefficient of the runner to experimental values 
of previous studies (i.e. validation).

\subsubsection*{Grid convergence study}
\label{Sec:A1}

\begin{figure}[htbp]
\begin{centering}
\vspace{0.2cm}
\begin{overpic}[width=0.50\textwidth]{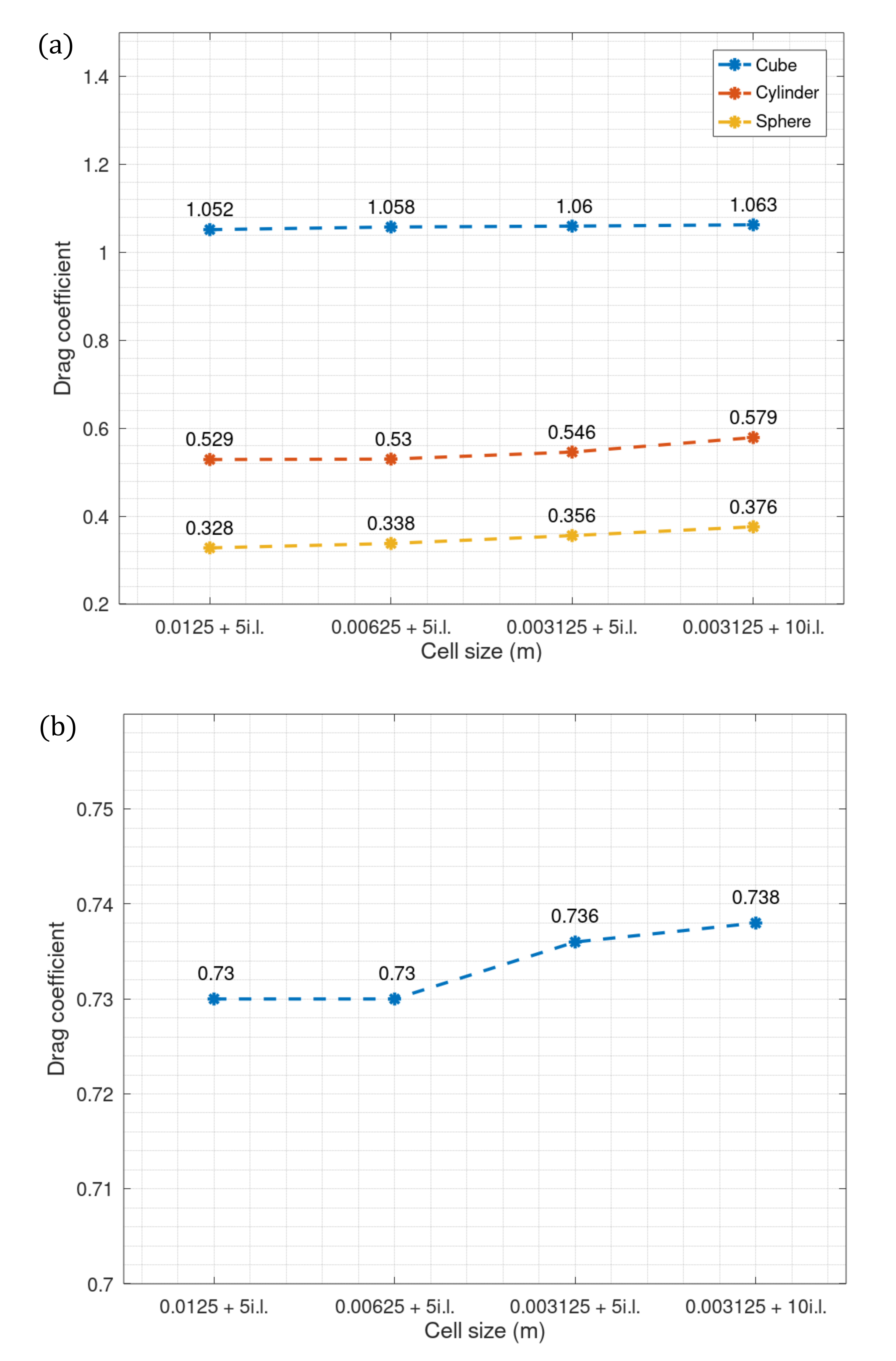}
\end{overpic}
\par
\end{centering}
\caption{Mesh convergence demonstrated by the final four refinement stages.
The drag coefficient of typical bluff bodies (a) and of the runner geometry used for this
study (b) are shown (i.l. stands for inflation layers).}
\label{Fig:A1-1} 
\end{figure}

The average cell size of the mesh at the finest grid level outside the prism layers
has been refined by a factor of two to obtain the target cell sizes at the runner's surface as illustrated in Fig. \ref{Fig:A1-1}.
All grids contain $5$ inflation layers (i.e. prism layers) at the object surfaces, 
except the final one, which has $10$ inflation layers.
It can be seen in Fig. \ref{Fig:A1-1} that for both the drag of the considered bluff bodies
as well as for the applied runner geometry the resulting changes in the final refinement stages are small.
Following \citet{roache1997quantification}, we can calculate the discretization error 
of two meshes with refinement ratio $r=h_{2}/h_{1}=0.5$ and with cell sizes $h_{1}$ and $h_{2}$
by using the difference $\phi_{2}-\phi_{1}$ between the two solutions of the target quantity, as
\begin{eqnarray}
E_{\phi,1} & = & \frac{\phi_{2}-\phi_{1}}{1-r^{p}},  \label{Eq:2D-1} \\
E_{\phi,2} & = & r^{p} \left( \frac{\phi_{2}-\phi_{1}}{1-r^{p}} \right),  \label{Eq:2D-2} 
\end{eqnarray}
with the order of the numerical scheme $p=2$.
Using this procedure, we obtain \\
$E_{\phi,1}=[E_{\phi,1,cube},E_{\phi,1,cylinder},E_{\phi,1,sphere}]=[0.0027,0.0213,0.0240]$ and 
$E_{\phi,2}=[E_{\phi,2,cube},E_{\phi,2,cylinder},E_{\phi,2,sphere}]=[0.0007,0.0053,0.0060]$ for the cell sizes 
$\SI{0.00625}{\meter}$ and $\SI{0.00312}{\meter}$, both with five inflation layers.
For the applied runner geometry the discretization errors are computed as 
$E_{\phi,1,runner}=0.0027$ and $E_{\phi,2,runner}=0.0007$.
Already with the coarser mesh the error is consistently below $3\%$ for all considered bluff bodies, 
and with the fine mesh it is below $1\%$.
Apart from the discretization error there is also a linearization error inherent to 
all CFD simulations, which comes from the linearization of the governing equations
through numerical schemes. 
Minimization of the linearization errors of the simulations is ensured through monitoring of 
the residuals of all flow variables and the convergence of the aerodynamic forces.
We apply a second-order scheme for increased accuracy
as well as tolerance values for all residuals of $\mathcal{O}(10^{-6})$.

\subsubsection*{Drag coefficients of general bluff bodies}
\label{Sec:A2}

Using the numerical results for the drag coefficients of the bluff bodies,
a comparison is done with experimental data from literature.
The results are summarized in Tab. \ref{Tab:A1-1}. 
\begin{table}[htbp]
\caption{Numerical values from this study and experimental values from literature
for typical bluff-body drag coefficients $C_{D}$ at a considered free-stream flow velocity of 
$U_{\infty} = \SI{18}{\kilo\meter\per\hour}$ ($\SI{5}{\meter\per\second}$) and a Reynolds number of $Re \approx (1.5\text{-}1.6) \times 10^{5}$.}
\centering
\begin{tabular}{@{}llll@{}}
\toprule
Object & $C_{D}$ (num.) & $C_{D}$ (exp.) \\
\midrule
Cube & $1.06$ & $1.05$ \citep{hoerner1951aerodynamic} \\
Cylinder & $0.58$ & $0.60$ \citep{prosser2015aerodynamics} \\
Sphere & $0.38$ & $0.40$ \citep{munsonfundamentals} \\
\bottomrule
\end{tabular}
\label{Tab:A1-1}
\end{table}
It can be seen that there is an overall good agreement between 
the numerical results and literature, despite the fact that the considered
Reynolds number is close to the drag crisis for the sphere and the
cylinder in crossflow. 
In this range there is a transition from laminar to turbulent boundary
layers, which leads to a sudden drop in the viscous drag due to delayed
flow separation. It is therefore crucial to properly resolve the 
flow field across the boundary layer, which is achieved here through 
insertion of a sufficient number of prism layers.

\subsubsection*{Drag coefficients of runners}
\label{Sec:A3}

In the final part of the validation the runner's drag is compared to the 
results of several experimental studies on runners.
\begin{table}[htbp]
\caption{Comparison of values from literature in chronological order
for the drag coefficient $C_{D}$ of humans at various activities.}
\centering
\begin{tabular}{@{}llll@{}}
\toprule
Authors & Activity & $C_{D}$ \\
\midrule
\citet{walpert1989aerodynamics} & Running & $0.79$ \\ \hdashline
\citet{davies1980effects} & Running & $0.82$-$0.91$ \\ \hdashline
\citet{pugh1971influence} & Running & $0.8$ \\ 
~ & Walking & $0.7$ \\ \hdashline
\citet{hoerner1965fluid} & Standing & $1.0$-$1.3$ (clothed) \\
~ & ~ & $0.9$-$1.2$ (nude) \\ \hdashline
\citet{hill1928air} & Running & $0.9$ \\
~ & Standing & $0.98$ \\
\bottomrule
\end{tabular}
\label{Tab:A1-2}
\end{table}
From the data in Tab. \ref{Tab:A1-2} it is obvious that a spread 
of the drag coefficient values exists with respect to the different activities,
but also for running itself. It proves the difficulty of establishing well-defined 
drag coefficients for humans experimentally.
However, there is a clear reduction of the drag coefficient from
standing to walking and running.
This has been argued to be due to the limbs being partially lifted in the 
air leading to a temporary reduction of frontal area and the subject being less affected by the air resistance over time.
Furthermore, clothing has a significant impact as well, leading to increased
drag and possibly earlier flow separation, especially if loose \citep{norstrud2009sport}.
The quantitative effect of surface properties on running aerodynamics has been assessed by \citet{kyle1986effect}, who estimate
the increase of drag by loose clothing to be up to $4.2\%$
and by hair to be $4$-$6\%$, depending on length.
In the current study, we obtain a drag coefficient of the considered runner
of $0.74$ at a comparable Reynolds number, which is close to the 
values established by \citet{pugh1971influence}, \citet{davies1980effects}, and \citet{walpert1989aerodynamics}.
One reason, why the drag coefficient computed in our case is slightly lower
than the values from literature given in Tab. \ref{Tab:A1-2} might be the smooth geometry, allowing
for delayed flow separation and thus reduced form drag on the runner.
This is an effect that would not occur in a wind tunnel setup involving 
real runners or their scaled models due to the aforementioned conditions of surface roughness, 
hair, and clothing.


\begin{thebibliography}{59}
\expandafter\ifx\csname natexlab\endcsname\relax\def\natexlab#1{#1}\fi
\providecommand{\url}[1]{\texttt{#1}}
\providecommand{\href}[2]{#2}
\providecommand{\path}[1]{#1}
\providecommand{\DOIprefix}{doi:}
\providecommand{\ArXivprefix}{arXiv:}
\providecommand{\URLprefix}{URL: }
\providecommand{\Pubmedprefix}{pmid:}
\providecommand{\doi}[1]{\href{http://dx.doi.org/#1}{\path{#1}}}
\providecommand{\Pubmed}[1]{\href{pmid:#1}{\path{#1}}}
\providecommand{\bibinfo}[2]{#2}
\ifx\xfnm\undefined \def\xfnm[#1]{\unskip,\space#1}\fi
\bibitem[{{American Institute of Aeronautics and
  Astronautics}(1998)}]{american1998aiaa}
\bibinfo{author}{{American Institute of Aeronautics and Astronautics}\xfnm[]}.
\newblock \bibinfo{title}{AIAA Guide for the Verification and Validation of
  Computational Fluid Dynamics Simulations}, \bibinfo{year}{1998}.
\bibitem[{Barnes and Kilding(2015)}]{barnes2015running}
\bibinfo{author}{Barnes\xfnm[ K.R.]}, \bibinfo{author}{Kilding\xfnm[ A.E.]}.
\newblock \bibinfo{title}{Running economy: Measurement, norms, and determining
  factors}.
\newblock \bibinfo{journal}{Sports Medicine}
  \bibinfo{year}{2015};\bibinfo{volume}{1}(\bibinfo{number}{1}):\bibinfo{pages}{1--15}.
\bibitem[{Bassett~Jr et~al.(1991)Bassett~Jr, Flohr, Duey, Howley and
  Pein}]{bassett1991metabolic}
\bibinfo{author}{Bassett~Jr\xfnm[ D.R.]}, \bibinfo{author}{Flohr\xfnm[ J.]},
  \bibinfo{author}{Duey\xfnm[ W.J.]}, \bibinfo{author}{Howley\xfnm[ E.T.]},
  \bibinfo{author}{Pein\xfnm[ R.L.]}.
\newblock \bibinfo{title}{Metabolic responses to drafting during front crawl
  swimming.}
\newblock \bibinfo{journal}{Medicine and Science in Sports and Exercise}
  \bibinfo{year}{1991};\bibinfo{volume}{23}(\bibinfo{number}{6}):\bibinfo{pages}{744--747}.
\bibitem[{Batliner et~al.(2018)Batliner, Kipp, Grabowski, Kram and
  Byrnes}]{batliner2018does}
\bibinfo{author}{Batliner\xfnm[ M.E.]}, \bibinfo{author}{Kipp\xfnm[ S.]},
  \bibinfo{author}{Grabowski\xfnm[ A.M.]}, \bibinfo{author}{Kram\xfnm[ R.]},
  \bibinfo{author}{Byrnes\xfnm[ W.C.]}.
\newblock \bibinfo{title}{Does metabolic rate increase linearly with running
  speed in all distance runners?}
\newblock \bibinfo{journal}{Sports Medicine International Open}
  \bibinfo{year}{2018};\bibinfo{volume}{2}(\bibinfo{number}{1}):\bibinfo{pages}{E1}.
\bibitem[{Beaumont et~al.(2019)Beaumont, Bogard, Murer, Polidori, Madaci and
  Taiar}]{beaumont2019does}
\bibinfo{author}{Beaumont\xfnm[ F.]}, \bibinfo{author}{Bogard\xfnm[ F.]},
  \bibinfo{author}{Murer\xfnm[ S.]}, \bibinfo{author}{Polidori\xfnm[ G.]},
  \bibinfo{author}{Madaci\xfnm[ F.]}, \bibinfo{author}{Taiar\xfnm[ R.]}.
\newblock \bibinfo{title}{How does aerodynamics influence physiological
  responses in middle-distance running drafting?}
\newblock \bibinfo{journal}{Mathematical Modelling of Engineering Problems}
  \bibinfo{year}{2019};\bibinfo{volume}{6}:\bibinfo{pages}{129--135}.
\bibitem[{Blocken et~al.(2013)Blocken, Defraeye, Koninckx, Carmeliet and
  Hespel}]{blocken2013cfd}
\bibinfo{author}{Blocken\xfnm[ B.]}, \bibinfo{author}{Defraeye\xfnm[ T.]},
  \bibinfo{author}{Koninckx\xfnm[ E.]}, \bibinfo{author}{Carmeliet\xfnm[ J.]},
  \bibinfo{author}{Hespel\xfnm[ P.]}.
\newblock \bibinfo{title}{{CFD} simulations of the aerodynamic drag of two
  drafting cyclists}.
\newblock \bibinfo{journal}{Computers \& Fluids}
  \bibinfo{year}{2013};\bibinfo{volume}{71}:\bibinfo{pages}{435--445}.
\bibitem[{Blocken et~al.(2018)Blocken, Toparlar, van Druenen and
  Andrianne}]{blocken2018aerodynamic}
\bibinfo{author}{Blocken\xfnm[ B.]}, \bibinfo{author}{Toparlar\xfnm[ Y.]},
  \bibinfo{author}{van Druenen\xfnm[ T.]}, \bibinfo{author}{Andrianne\xfnm[
  T.]}.
\newblock \bibinfo{title}{Aerodynamic drag in cycling team time trials}.
\newblock \bibinfo{journal}{Journal of Wind Engineering and Industrial
  Aerodynamics}
  \bibinfo{year}{2018};\bibinfo{volume}{182}:\bibinfo{pages}{128--145}.
\bibitem[{Cavagna et~al.(1968)Cavagna, Dusman and
  Margaria}]{cavagna1968positive}
\bibinfo{author}{Cavagna\xfnm[ G.]}, \bibinfo{author}{Dusman\xfnm[ B.]},
  \bibinfo{author}{Margaria\xfnm[ R.]}.
\newblock \bibinfo{title}{Positive work done by a previously stretched muscle.}
\newblock \bibinfo{journal}{Journal of Applied Physiology}
  \bibinfo{year}{1968};\bibinfo{volume}{24}(\bibinfo{number}{1}):\bibinfo{pages}{21--32}.
\bibitem[{Cavagna and Kaneko(1977)}]{cavagna1977mechanical}
\bibinfo{author}{Cavagna\xfnm[ G.]}, \bibinfo{author}{Kaneko\xfnm[ M.]}.
\newblock \bibinfo{title}{Mechanical work and efficiency in level walking and
  running}.
\newblock \bibinfo{journal}{Journal of Physiology}
  \bibinfo{year}{1977};\bibinfo{volume}{268}(\bibinfo{number}{2}):\bibinfo{pages}{467--481}.
\bibitem[{Chatard and Wilson(2003)}]{chatard2003drafting}
\bibinfo{author}{Chatard\xfnm[ J.C.]}, \bibinfo{author}{Wilson\xfnm[ B.]}.
\newblock \bibinfo{title}{Drafting distance in swimming}.
\newblock \bibinfo{journal}{Medicine and Science in Sports and Exercise}
  \bibinfo{year}{2003};\bibinfo{volume}{35}(\bibinfo{number}{7}):\bibinfo{pages}{1176--1181}.
\bibitem[{Colaianni et~al.(2014)Colaianni, Zollh{\"o}fer, S{\"u}{\ss}muth,
  Seider and Greiner}]{colaianni2014pose}
\bibinfo{author}{Colaianni\xfnm[ M.]}, \bibinfo{author}{Zollh{\"o}fer\xfnm[
  M.]}, \bibinfo{author}{S{\"u}{\ss}muth\xfnm[ J.]},
  \bibinfo{author}{Seider\xfnm[ B.]}, \bibinfo{author}{Greiner\xfnm[ G.]}.
\newblock \bibinfo{title}{A pose invariant statistical shape model for human
  bodies}.
\newblock In: \bibinfo{booktitle}{Proceedings of the 5th International
  Conference on 3D Body Scanning Technologies}. \bibinfo{year}{2014}. p.
  \bibinfo{pages}{327--336}.
\bibitem[{Crouch et~al.(2016)Crouch, Burton, Thompson, Brown and
  Sheridan}]{crouch2016dynamic}
\bibinfo{author}{Crouch\xfnm[ T.N.]}, \bibinfo{author}{Burton\xfnm[ D.]},
  \bibinfo{author}{Thompson\xfnm[ M.C.]}, \bibinfo{author}{Brown\xfnm[ N.A.]},
  \bibinfo{author}{Sheridan\xfnm[ J.]}.
\newblock \bibinfo{title}{Dynamic leg-motion and its effect on the aerodynamic
  performance of cyclists}.
\newblock \bibinfo{journal}{Journal of Fluids and Structures}
  \bibinfo{year}{2016};\bibinfo{volume}{65}:\bibinfo{pages}{121--137}.
\bibitem[{Davies(1980)}]{davies1980effects}
\bibinfo{author}{Davies\xfnm[ C.]}.
\newblock \bibinfo{title}{Effects of wind assistance and resistance on the
  forward motion of a runner}.
\newblock \bibinfo{journal}{Journal of Applied Physiology}
  \bibinfo{year}{1980};\bibinfo{volume}{48}(\bibinfo{number}{4}):\bibinfo{pages}{702--709}.
\bibitem[{Defraeye et~al.(2010)Defraeye, Blocken, Koninckx, Hespel and
  Carmeliet}]{defraeye2010computational}
\bibinfo{author}{Defraeye\xfnm[ T.]}, \bibinfo{author}{Blocken\xfnm[ B.]},
  \bibinfo{author}{Koninckx\xfnm[ E.]}, \bibinfo{author}{Hespel\xfnm[ P.]},
  \bibinfo{author}{Carmeliet\xfnm[ J.]}.
\newblock \bibinfo{title}{Computational fluid dynamics analysis of cyclist
  aerodynamics: Performance of different turbulence-modelling and
  boundary-layer modelling approaches}.
\newblock \bibinfo{journal}{Journal of Biomechanics}
  \bibinfo{year}{2010};\bibinfo{volume}{43}(\bibinfo{number}{12}):\bibinfo{pages}{2281--2287}.
\bibitem[{Dickinson(1929)}]{dickinson1929efficiency}
\bibinfo{author}{Dickinson\xfnm[ S.]}.
\newblock \bibinfo{title}{The efficiency of bicycle-pedalling, as affected by
  speed and load}.
\newblock \bibinfo{journal}{Journal of Physiology}
  \bibinfo{year}{1929};\bibinfo{volume}{67}(\bibinfo{number}{3}):\bibinfo{pages}{242--255}.
\bibitem[{Edwards and Byrnes(2007)}]{edwards2007aerodynamic}
\bibinfo{author}{Edwards\xfnm[ A.G.]}, \bibinfo{author}{Byrnes\xfnm[ W.C.]}.
\newblock \bibinfo{title}{Aerodynamic characteristics as determinants of the
  drafting effect in cycling.}
\newblock \bibinfo{journal}{Medicine and Science in Sports and Exercise}
  \bibinfo{year}{2007};\bibinfo{volume}{39}(\bibinfo{number}{1}):\bibinfo{pages}{170--176}.
\bibitem[{Ettema and Lor{\aa}s(2009)}]{ettema2009efficiency}
\bibinfo{author}{Ettema\xfnm[ G.]}, \bibinfo{author}{Lor{\aa}s\xfnm[ H.W.]}.
\newblock \bibinfo{title}{Efficiency in cycling: A review}.
\newblock \bibinfo{journal}{European Journal of Applied Physiology}
  \bibinfo{year}{2009};\bibinfo{volume}{106}(\bibinfo{number}{1}):\bibinfo{pages}{1--14}.
\bibitem[{Fitton et~al.(2018)Fitton, Caddy and Symons}]{fitton2018impact}
\bibinfo{author}{Fitton\xfnm[ B.]}, \bibinfo{author}{Caddy\xfnm[ O.]},
  \bibinfo{author}{Symons\xfnm[ D.]}.
\newblock \bibinfo{title}{The impact of relative athlete characteristics on the
  drag reductions caused by drafting when cycling in a velodrome}.
\newblock \bibinfo{journal}{Proceedings of the Institution of Mechanical
  Engineers, Part P: Journal of Sports Engineering and Technology}
  \bibinfo{year}{2018};\bibinfo{volume}{232}(\bibinfo{number}{1}):\bibinfo{pages}{39--49}.
\bibitem[{Fukunaga et~al.(1980)Fukunaga, Matsuo, Yuasa, Fujimatsu and
  Asahina}]{fukunaga1980effect}
\bibinfo{author}{Fukunaga\xfnm[ T.]}, \bibinfo{author}{Matsuo\xfnm[ A.]},
  \bibinfo{author}{Yuasa\xfnm[ K.]}, \bibinfo{author}{Fujimatsu\xfnm[ H.]},
  \bibinfo{author}{Asahina\xfnm[ K.]}.
\newblock \bibinfo{title}{Effect of running velocity on external mechanical
  power output}.
\newblock \bibinfo{journal}{Ergonomics}
  \bibinfo{year}{1980};\bibinfo{volume}{23}(\bibinfo{number}{2}):\bibinfo{pages}{123--136}.
\bibitem[{Fuss(2018)}]{fuss2018slipstreaming}
\bibinfo{author}{Fuss\xfnm[ F.K.]}.
\newblock \bibinfo{title}{Slipstreaming in gravity powered sports: Application
  to racing strategy in ski cross}.
\newblock \bibinfo{journal}{Frontiers in Physiology}
  \bibinfo{year}{2018};\bibinfo{volume}{9}:\bibinfo{pages}{1032}.
\bibitem[{Gray et~al.(2020)Gray, Andrews, Waldron and Jenkins}]{gray2020model}
\bibinfo{author}{Gray\xfnm[ A.]}, \bibinfo{author}{Andrews\xfnm[ M.]},
  \bibinfo{author}{Waldron\xfnm[ M.]}, \bibinfo{author}{Jenkins\xfnm[ D.]}.
\newblock \bibinfo{title}{A model for calculating the mechanical demands of
  overground running}.
\newblock \bibinfo{journal}{Sports Biomechanics}
  \bibinfo{year}{2020};:\bibinfo{pages}{1--22}.
\bibitem[{Hellsten(1998)}]{hellsten1998some}
\bibinfo{author}{Hellsten\xfnm[ A.]}.
\newblock \bibinfo{title}{Some improvements in {M}enter's k-$\omega$ {SST}
  turbulence model}.
\newblock In: \bibinfo{booktitle}{29th AIAA, Fluid Dynamics Conference}.
  \bibinfo{year}{1998}. p. \bibinfo{pages}{2554}.
\bibitem[{Hill(1928)}]{hill1928air}
\bibinfo{author}{Hill\xfnm[ A.V.]}.
\newblock \bibinfo{title}{The air-resistance to a runner}.
\newblock \bibinfo{journal}{Proceedings of the Royal Society of London Series
  B, Containing Papers of a Biological Character}
  \bibinfo{year}{1928};\bibinfo{volume}{102}(\bibinfo{number}{718}):\bibinfo{pages}{380--385}.
\bibitem[{Hill(1964)}]{hill1964efficiency}
\bibinfo{author}{Hill\xfnm[ A.V.]}.
\newblock \bibinfo{title}{The efficiency of mechanical power development during
  muscular shortening and its relation to load}.
\newblock \bibinfo{journal}{Proceedings of the Royal Society of London Series B
  Biological Sciences}
  \bibinfo{year}{1964};\bibinfo{volume}{159}(\bibinfo{number}{975}):\bibinfo{pages}{319--324}.
\bibitem[{Hoerner(1951)}]{hoerner1951aerodynamic}
\bibinfo{author}{Hoerner\xfnm[ S.F.]}.
\newblock \bibinfo{title}{Aerodynamic Drag: Practical Data on Aerodynamic Drag
  Evaluated and Presented}.
\newblock \bibinfo{publisher}{Otterbein Press}, \bibinfo{year}{1951}.
\bibitem[{Hoerner(1965)}]{hoerner1965fluid}
\bibinfo{author}{Hoerner\xfnm[ S.F.]}.
\newblock \bibinfo{title}{Fluid-Dynamic Drag: Theoretical, experimental and
  statistical information}.
\newblock \bibinfo{publisher}{SF Hoerner Fluid Dynamics}, \bibinfo{year}{1965}.
\bibitem[{Hoogkamer et~al.(2018)Hoogkamer, Snyder and
  Arellano}]{hoogkamer2018modeling}
\bibinfo{author}{Hoogkamer\xfnm[ W.]}, \bibinfo{author}{Snyder\xfnm[ K.L.]},
  \bibinfo{author}{Arellano\xfnm[ C.J.]}.
\newblock \bibinfo{title}{Modeling the benefits of cooperative drafting: Is
  there an optimal strategy to facilitate a sub-2-hour marathon performance?}
\newblock \bibinfo{journal}{Sports Medicine}
  \bibinfo{year}{2018};\bibinfo{volume}{48}(\bibinfo{number}{12}):\bibinfo{pages}{2859--2867}.
\bibitem[{Hunter et~al.(2015)Hunter, McCarthy, Carter, Bamman, Gaddy, Fisher,
  Katsoulis, Plaisance and Newcomer}]{hunter2015muscle}
\bibinfo{author}{Hunter\xfnm[ G.R.]}, \bibinfo{author}{McCarthy\xfnm[ J.P.]},
  \bibinfo{author}{Carter\xfnm[ S.J.]}, \bibinfo{author}{Bamman\xfnm[ M.M.]},
  \bibinfo{author}{Gaddy\xfnm[ E.S.]}, \bibinfo{author}{Fisher\xfnm[ G.]},
  \bibinfo{author}{Katsoulis\xfnm[ K.]}, \bibinfo{author}{Plaisance\xfnm[
  E.P.]}, \bibinfo{author}{Newcomer\xfnm[ B.R.]}.
\newblock \bibinfo{title}{Muscle fiber type, achilles tendon length,
  potentiation, and running economy}.
\newblock \bibinfo{journal}{Journal of Strength \& Conditioning Research}
  \bibinfo{year}{2015};\bibinfo{volume}{29}(\bibinfo{number}{5}):\bibinfo{pages}{1302--1309}.
\bibitem[{Inoue et~al.(2016)Inoue, Okayama, Teraoka, Maeno and
  Hirata}]{inoue2016wind}
\bibinfo{author}{Inoue\xfnm[ T.]}, \bibinfo{author}{Okayama\xfnm[ T.]},
  \bibinfo{author}{Teraoka\xfnm[ T.]}, \bibinfo{author}{Maeno\xfnm[ S.]},
  \bibinfo{author}{Hirata\xfnm[ K.]}.
\newblock \bibinfo{title}{Wind-tunnel experiment on aerodynamic characteristics
  of a runner using a moving-belt system}.
\newblock \bibinfo{journal}{Cogent Engineering}
  \bibinfo{year}{2016};\bibinfo{volume}{3}(\bibinfo{number}{1}):\bibinfo{pages}{1231389}.
\bibitem[{Katz(1995)}]{katz1995race}
\bibinfo{author}{Katz\xfnm[ J.]}.
\newblock \bibinfo{title}{Race Car Aerodynamics: Designing for Speed}.
\newblock \bibinfo{publisher}{Bentley Publishers}, \bibinfo{year}{1995}.
\bibitem[{Kipp(2017)}]{kipp2017does}
\bibinfo{author}{Kipp\xfnm[ S.]}.
\newblock \bibinfo{title}{Why does metabolic rate increase curvilinearly with
  running velocity?}
\newblock Master's thesis; University of Colorado, Boulder, Colorado;
  \bibinfo{year}{2017}.
\bibitem[{Kipp et~al.(2019)Kipp, Kram and Hoogkamer}]{kipp2019extrapolating}
\bibinfo{author}{Kipp\xfnm[ S.]}, \bibinfo{author}{Kram\xfnm[ R.]},
  \bibinfo{author}{Hoogkamer\xfnm[ W.]}.
\newblock \bibinfo{title}{Extrapolating metabolic savings in running:
  Implications for performance predictions}.
\newblock \bibinfo{journal}{Frontiers in Physiology}
  \bibinfo{year}{2019};\bibinfo{volume}{10}:\bibinfo{pages}{79}.
\bibitem[{Komi(2000)}]{komi2000stretch}
\bibinfo{author}{Komi\xfnm[ P.V.]}.
\newblock \bibinfo{title}{Stretch-shortening cycle: A powerful model to study
  normal and fatigued muscle}.
\newblock \bibinfo{journal}{Journal of Biomechanics}
  \bibinfo{year}{2000};\bibinfo{volume}{33}(\bibinfo{number}{10}):\bibinfo{pages}{1197--1206}.
\bibitem[{Komi and Bosco(1978)}]{komi1978utilization}
\bibinfo{author}{Komi\xfnm[ P.V.]}, \bibinfo{author}{Bosco\xfnm[ C.]}.
\newblock \bibinfo{title}{Utilization of stored elastic energy in leg extensor
  muscles by men and women.}
\newblock \bibinfo{journal}{Medicine and Science in Sports}
  \bibinfo{year}{1978};\bibinfo{volume}{10}(\bibinfo{number}{4}):\bibinfo{pages}{261--265}.
\bibitem[{Kyle and Caiozzo(1986)}]{kyle1986effect}
\bibinfo{author}{Kyle\xfnm[ C.R.]}, \bibinfo{author}{Caiozzo\xfnm[ V.J.]}.
\newblock \bibinfo{title}{The effect of athletic clothing aerodynamics upon
  running speed.}
\newblock \bibinfo{journal}{Medicine and Science in Sports and Exercise}
  \bibinfo{year}{1986};\bibinfo{volume}{18}(\bibinfo{number}{5}):\bibinfo{pages}{509--515}.
\bibitem[{Levenberg(1944)}]{levenberg1944method}
\bibinfo{author}{Levenberg\xfnm[ K.]}.
\newblock \bibinfo{title}{A method for the solution of certain non-linear
  problems in least squares}.
\newblock \bibinfo{journal}{Quarterly of Applied Mathematics}
  \bibinfo{year}{1944};\bibinfo{volume}{2}(\bibinfo{number}{2}):\bibinfo{pages}{164--168}.
\bibitem[{Malizia and Blocken(2020)}]{malizia2020bicycle}
\bibinfo{author}{Malizia\xfnm[ F.]}, \bibinfo{author}{Blocken\xfnm[ B.]}.
\newblock \bibinfo{title}{Bicycle aerodynamics: History, state-of-the-art and
  future perspectives}.
\newblock \bibinfo{journal}{Journal of Wind Engineering and Industrial
  Aerodynamics}
  \bibinfo{year}{2020};\bibinfo{volume}{200}:\bibinfo{pages}{104134}.
\bibitem[{Marquardt(1963)}]{marquardt1963algorithm}
\bibinfo{author}{Marquardt\xfnm[ D.W.]}.
\newblock \bibinfo{title}{An algorithm for least-squares estimation of
  nonlinear parameters}.
\newblock \bibinfo{journal}{Journal of the Society for Industrial and Applied
  Mathematics}
  \bibinfo{year}{1963};\bibinfo{volume}{11}(\bibinfo{number}{2}):\bibinfo{pages}{431--441}.
\bibitem[{Meile et~al.(2006)Meile, Reisenberger, Mayer, Schm{\"o}lzer,
  M{\"u}ller and Brenn}]{meile2006aerodynamics}
\bibinfo{author}{Meile\xfnm[ W.]}, \bibinfo{author}{Reisenberger\xfnm[ E.]},
  \bibinfo{author}{Mayer\xfnm[ M.]}, \bibinfo{author}{Schm{\"o}lzer\xfnm[ B.]},
  \bibinfo{author}{M{\"u}ller\xfnm[ W.]}, \bibinfo{author}{Brenn\xfnm[ G.]}.
\newblock \bibinfo{title}{Aerodynamics of ski jumping: Experiments and {CFD}
  simulations}.
\newblock \bibinfo{journal}{Experiments in Fluids}
  \bibinfo{year}{2006};\bibinfo{volume}{41}(\bibinfo{number}{6}):\bibinfo{pages}{949--964}.
\bibitem[{Menter(1992)}]{menter1992improved}
\bibinfo{author}{Menter\xfnm[ F.R.]}.
\newblock \bibinfo{title}{Improved two-equation k-$\omega$-turbulence models
  for aerodynamic flows}.
\newblock \bibinfo{journal}{National Aeronautics and Space Administration, Ames
  Research Center, Moffett Field, CA} \bibinfo{year}{1992};.
\bibitem[{Menter(1994)}]{menter1994two}
\bibinfo{author}{Menter\xfnm[ F.R.]}.
\newblock \bibinfo{title}{Two-equation eddy-viscosity turbulence models for
  engineering applications}.
\newblock \bibinfo{journal}{AIAA Journal}
  \bibinfo{year}{1994};\bibinfo{volume}{32}(\bibinfo{number}{8}):\bibinfo{pages}{1598--1605}.
\bibitem[{Minetti et~al.(2001)Minetti, Pinkerton and
  Zamparo}]{minetti2001bipedalism}
\bibinfo{author}{Minetti\xfnm[ A.E.]}, \bibinfo{author}{Pinkerton\xfnm[ J.]},
  \bibinfo{author}{Zamparo\xfnm[ P.]}.
\newblock \bibinfo{title}{From bipedalism to bicyclism: Evolution in energetics
  and biomechanics of historic bicycles}.
\newblock \bibinfo{journal}{Proceedings of the Royal Society of London Series
  B: Biological Sciences}
  \bibinfo{year}{2001};\bibinfo{volume}{268}(\bibinfo{number}{1474}):\bibinfo{pages}{1351--1360}.
\bibitem[{Munson et~al.(1990)Munson, Young and Okiishi}]{munsonfundamentals}
\bibinfo{author}{Munson\xfnm[ B.R.]}, \bibinfo{author}{Young\xfnm[ D.F.]},
  \bibinfo{author}{Okiishi\xfnm[ T.H.]}.
\newblock \bibinfo{title}{Fundamentals of Fluid Mechanics}.
\newblock \bibinfo{publisher}{John Wiley \& Sons}, \bibinfo{year}{1990}.
\bibitem[{N{\o}rstrud(2009)}]{norstrud2009sport}
\bibinfo{author}{N{\o}rstrud\xfnm[ H.]}.
\newblock \bibinfo{title}{Sport Aerodynamics}.
\newblock \bibinfo{publisher}{Springer}, \bibinfo{year}{2009}.
\bibitem[{Pavei et~al.(2019)Pavei, Zamparo, Fujii, Otsu, Numazu, Minetti and
  Monte}]{pavei2019comprehensive}
\bibinfo{author}{Pavei\xfnm[ G.]}, \bibinfo{author}{Zamparo\xfnm[ P.]},
  \bibinfo{author}{Fujii\xfnm[ N.]}, \bibinfo{author}{Otsu\xfnm[ T.]},
  \bibinfo{author}{Numazu\xfnm[ N.]}, \bibinfo{author}{Minetti\xfnm[ A.E.]},
  \bibinfo{author}{Monte\xfnm[ A.]}.
\newblock \bibinfo{title}{Comprehensive mechanical power analysis in sprint
  running acceleration}.
\newblock \bibinfo{journal}{Scandinavian Journal of Medicine \& Science in
  Sports}
  \bibinfo{year}{2019};\bibinfo{volume}{29}(\bibinfo{number}{12}):\bibinfo{pages}{1892--1900}.
\bibitem[{Polidori et~al.(2020)Polidori, Legrand, Bogard, Madaci and
  Beaumont}]{polidori2020numerical}
\bibinfo{author}{Polidori\xfnm[ G.]}, \bibinfo{author}{Legrand\xfnm[ F.]},
  \bibinfo{author}{Bogard\xfnm[ F.]}, \bibinfo{author}{Madaci\xfnm[ F.]},
  \bibinfo{author}{Beaumont\xfnm[ F.]}.
\newblock \bibinfo{title}{Numerical investigation of the impact of {K}enenisa
  {B}ekele's cooperative drafting strategy on its running power during the 2019
  {B}erlin marathon}.
\newblock \bibinfo{journal}{Journal of Biomechanics}
  \bibinfo{year}{2020};\bibinfo{volume}{107}:\bibinfo{pages}{109854}.
\bibitem[{Prosser and Smith(2015)}]{prosser2015aerodynamics}
\bibinfo{author}{Prosser\xfnm[ D.]}, \bibinfo{author}{Smith\xfnm[ M.]}.
\newblock \bibinfo{title}{Aerodynamics of finite cylinders in quasi-steady
  flow}.
\newblock In: \bibinfo{booktitle}{53rd AIAA Aerospace Sciences Meeting}.
  \bibinfo{year}{2015}. p. \bibinfo{pages}{1931}.
\bibitem[{Pugh(1970)}]{pugh1970oxygen}
\bibinfo{author}{Pugh\xfnm[ L.G.E.]}.
\newblock \bibinfo{title}{Oxygen intake in track and treadmill running with
  observations on the effect of air resistance}.
\newblock \bibinfo{journal}{Journal of Physiology}
  \bibinfo{year}{1970};\bibinfo{volume}{207}(\bibinfo{number}{3}):\bibinfo{pages}{823--835}.
\bibitem[{Pugh(1971)}]{pugh1971influence}
\bibinfo{author}{Pugh\xfnm[ L.G.E.]}.
\newblock \bibinfo{title}{The influence of wind resistance in running and
  walking and the mechanical efficiency of work against horizontal or vertical
  forces}.
\newblock \bibinfo{journal}{Journal of Physiology}
  \bibinfo{year}{1971};\bibinfo{volume}{213}(\bibinfo{number}{2}):\bibinfo{pages}{255--276}.
\bibitem[{Roache(1997)}]{roache1997quantification}
\bibinfo{author}{Roache\xfnm[ P.J.]}.
\newblock \bibinfo{title}{Quantification of uncertainty in computational fluid
  dynamics}.
\newblock \bibinfo{journal}{Annual Review of Fluid Mechanics}
  \bibinfo{year}{1997};\bibinfo{volume}{29}(\bibinfo{number}{1}):\bibinfo{pages}{123--160}.
\bibitem[{Roache(1998)}]{roache1998verification}
\bibinfo{author}{Roache\xfnm[ P.J.]}.
\newblock \bibinfo{title}{Verification and Validation in Computational Science
  and Engineering}.
\newblock \bibinfo{publisher}{Hermosa Publishers}, \bibinfo{year}{1998}.
\bibitem[{Romberg et~al.(1971)Romberg, Chianese and
  Lajoie}]{romberg1971aerodynamics}
\bibinfo{author}{Romberg\xfnm[ G.]}, \bibinfo{author}{Chianese\xfnm[ F.]},
  \bibinfo{author}{Lajoie\xfnm[ R.]}.
\newblock \bibinfo{title}{Aerodynamics of race cars in drafting and passing
  situations}.
\newblock \bibinfo{type}{Technical Report}; SAE; \bibinfo{year}{1971}.
\bibitem[{Rundell(1996)}]{rundell1996effects}
\bibinfo{author}{Rundell\xfnm[ K.W.]}.
\newblock \bibinfo{title}{Effects of drafting during short-track speed
  skating.}
\newblock \bibinfo{journal}{Medicine and Science in Sports and Exercise}
  \bibinfo{year}{1996};\bibinfo{volume}{28}(\bibinfo{number}{6}):\bibinfo{pages}{765--771}.
\bibitem[{Saibene and Minetti(2003)}]{saibene2003biomechanical}
\bibinfo{author}{Saibene\xfnm[ F.]}, \bibinfo{author}{Minetti\xfnm[ A.E.]}.
\newblock \bibinfo{title}{Biomechanical and physiological aspects of legged
  locomotion in humans}.
\newblock \bibinfo{journal}{European Journal of Applied Physiology}
  \bibinfo{year}{2003};\bibinfo{volume}{88}(\bibinfo{number}{4-5}):\bibinfo{pages}{297--316}.
\bibitem[{Saunders et~al.(2004)Saunders, Pyne, Telford and
  Hawley}]{saunders2004factors}
\bibinfo{author}{Saunders\xfnm[ P.U.]}, \bibinfo{author}{Pyne\xfnm[ D.B.]},
  \bibinfo{author}{Telford\xfnm[ R.D.]}, \bibinfo{author}{Hawley\xfnm[ J.A.]}.
\newblock \bibinfo{title}{Factors affecting running economy in trained distance
  runners}.
\newblock \bibinfo{journal}{Sports Medicine}
  \bibinfo{year}{2004};\bibinfo{volume}{34}(\bibinfo{number}{7}):\bibinfo{pages}{465--485}.
\bibitem[{Walpert and Kyle(1989)}]{walpert1989aerodynamics}
\bibinfo{author}{Walpert\xfnm[ R.A.]}, \bibinfo{author}{Kyle\xfnm[ C.R.]}.
\newblock \bibinfo{title}{Aerodynamics of the human body in sports}.
\newblock \bibinfo{journal}{Journal of Biomechanics}
  \bibinfo{year}{1989};\bibinfo{volume}{22}(\bibinfo{number}{10}):\bibinfo{pages}{1096}.
\bibitem[{Williams(2000)}]{williams2000dynamics}
\bibinfo{author}{Williams\xfnm[ K.R.]}.
\newblock \bibinfo{title}{Biomechanics in Sport}; \bibinfo{publisher}{Wiley
  Online Library}.
\newblock p. \bibinfo{pages}{161}.
\bibitem[{Williams and Cavanagh(1987)}]{williams1987relationship}
\bibinfo{author}{Williams\xfnm[ K.R.]}, \bibinfo{author}{Cavanagh\xfnm[ P.R.]}.
\newblock \bibinfo{title}{Relationship between distance running mechanics,
  running economy, and performance}.
\newblock \bibinfo{journal}{Journal of Applied Physiology}
  \bibinfo{year}{1987};\bibinfo{volume}{63}(\bibinfo{number}{3}):\bibinfo{pages}{1236--1245}.
\bibitem[{Winter(1979)}]{winter1979new}
\bibinfo{author}{Winter\xfnm[ D.A.]}.
\newblock \bibinfo{title}{A new definition of mechanical work done in human
  movement}.
\newblock \bibinfo{journal}{Journal of Applied Physiology}
  \bibinfo{year}{1979};\bibinfo{volume}{46}(\bibinfo{number}{1}):\bibinfo{pages}{79--83}.
\end{thebibliography}
\end{document}